\documentclass[11pt,oneside,letterpaper]{article}
\usepackage{amssymb}
\usepackage{amsmath}
\usepackage{graphicx}
\usepackage{setspace}
\usepackage{amsfonts}
\usepackage{leftidx}
\usepackage{relsize}
\usepackage{tikz-cd}
\usepackage{tikz}
\usetikzlibrary{decorations.markings}
\usepackage{pgfplots}
\usetikzlibrary{cd}
\usepackage{amsthm}
\usepackage{fancyhdr}
\usepackage{xcolor}
\usepackage{rotating}
\usepackage{comment}
\usepackage{color}
\usepackage{cite}
\usepackage{braket}
\usepackage{enumerate}
\definecolor{darkgreen}{rgb}{0,0.5,0}
\definecolor{darkblue}{rgb}{0,0,0.5}
\definecolor{darkred}{rgb}{0.5,0,0}
\usepackage[colorlinks=true,citecolor=darkgreen,linkcolor=darkred,urlcolor=darkblue]{hyperref}
\usepackage{url}
\usepackage{pdfsync}

\def\ben{\begin{equation}}
\def\een{\end{equation}}

\let\a=\alpha \let\b=\beta   
   
\let\l=\lambda \let\m=\mu    \let\r=v
\let\s=\sigma \let\t=\tau

\let\C=\Chi

\def\be{\begin{equation}}
\def\ee{\end{equation}}
\def\ba{\begin{array}}
\def\ea{\end{array}}

\def\dalemb#1#2{{\vbox{\hrule height .#2pt
       \hbox{\vrule width.#2pt height#1pt \kern#1pt
               \vrule width.#2pt}
       \hrule height.#2pt}}}

\newcommand{\bea}{\begin{eqnarray}}
\newcommand{\eea}{\end{eqnarray}}

\newcommand{\Tr}{{\rm Tr} }

\def\R{{{\mathbf{R}}}}
\def\C{{{\mathbf{C}}}}

\def\Z{{{\mathbf{Z}}}}

\let\tilde=\widetilde

\def\pt{{\rm pt}}

\def\im{\operatorname{im}}

\theoremstyle{plain}

\theoremstyle{definition}

\theoremstyle{remark}

\numberwithin{equation}{section}
\allowdisplaybreaks  

\begin{document}
\onehalfspacing

\vspace*{0.5cm}
\begin{center}
{ \LARGE \textsc{Classification of crystalline topological insulators through $K$-theory}} \\
\vspace*{1.7cm}

\vskip10mm

Luuk Stehouwer, $^\clubsuit$ \ \ Jan de Boer,$^{\spadesuit}$\ \ Jorrit Kruthoff,$^{\spadesuit}$\ \ Hessel Posthuma,$^\clubsuit$\ \ 
\vskip5mm

{$\spadesuit$  Institute for Theoretical Physics Amsterdam and Delta Institute for Theoretical Physics, University of Amsterdam, Science Park 904, 1098 XH Amsterdam, The Netherlands} \\
\vskip5mm
{$\clubsuit$  Korteweg-de Vries Institute for Mathematics, University of Amsterdam, Science Park 105-107, 1098 XG Amsterdam, The Netherlands} 

\vskip5mm

\tt{j.deboer@uva.nl, j.kruthoff@uva.nl, h.b.posthuma@uva.nl, luuk.stehouwer@gmail.com}


\end{center}
\vspace*{1.5cm}
\begin{abstract}
\noindent
Topological phases for free fermions in systems with crystal symmetry are classified by the topology of the valence band 
viewed as a vector bundle over the Brillouin zone. Additional symmetries, such as crystal symmetries which act non-trivially on the Brillouin zone, or time-reversal symmetry, endow the vector bundle with extra structure. These vector bundles
are classified by a suitable version of K-theory. While relatively easy to define, these K-theory groups are notoriously
hard to compute in explicit examples. In this paper we describe in detail how one can compute these K-theory groups starting
with a decomposition of the Brillouin zone in terms of simple submanifolds on which the symmetries act nicely. The main
mathematical tool is the Atiyah-Hirzebruch spectral sequence associated to such a decomposition, which will not only 
yield the explicit result for several crystal symmetries, but also sheds light on the origin of the topological invariants.
This extends results that have appeared in the literature so far. We also describe examples in which this approach 
fails to directly yield a conclusive answer, and discuss various open problems and directions for future research.


\end{abstract}

\newpage
\setcounter{page}{1}
\pagenumbering{arabic}

\tableofcontents
\setcounter{tocdepth}{2}


\section{Introduction}\label{sec:Intro}

Topological phases of matter form an interesting playground for both experimental and theoretical physics. These phases have the remarkable property to be resilient against external perturbations such as weak disorder or weak interactions. This emerges from a gap in the spectrum, either between the ground state and first excited state or around the Fermi level and most of the physics is contained in the states below the gap, i.e. in the degeneracy of the ground state or the occupied states below the Fermi energy. For systems consisting of free fermions moving in a crystal the occupied states form the valence bands and it is the topology of this part of the spectrum that makes a topological phase (of free fermions) topological. Using topological invariants one can capture this topology and, in fact, characterise the topological phase. The most well-known example of this is the Chern number that characterises the integer quantum Hall (IQHE) plateaus. 

We mentioned that a topological phase is resilient against external perturbations, but once these perturbations become too strong so that they cause the (band) gap to close, the topological phase is destroyed, either by becoming an ordinary insulator or by going to another topological phase. Deforming one topological phase into another allows us to understand how many topological phases there are and how to classify them. For free fermion systems with or without time-reversal symmetry and/or particle-hole symmetry, this program was initiated in \cite{PhysRevLett.95.016405, Kitaev2009, altlandzirnbauer}. In particular, Horava and Kitaev noticed an intricate relation between this classification and the classification of vector bundles using $K$-theory. It was not until the work of Freed and Moore \cite{Freed2013} that a complete proposal was formulated to classify topological phases of free fermions by including not only time-reversal or particle-hole symmetry, but also the crystal symmetries that these fermions experience. 

The proposal of Freed and Moore involves the computation of a suitable twisted and equivariant $K$-theory. It captures all topological invariants present for a given symmetry class and crystal. These invariants describe both global and local information of valence bands. Although we can treat these invariants in a unifying way, from a physics point of view, global and local invariants describe different aspects of the valence bands, which is why we will now discuss them separately. 

The local invariants \cite{Kruthoff1, Po2017} can be defined by carefully analyzing how crystal symmetries act on momentum space and the Hilbert space. They count the number of bands with a particular eigenvalue under the unbroken symmetry of the high-symmetry point in the Brillouin zone. For instance, when there is a fourfold rotation symmetry, the bands at the origin of the Brillouin zone are labelled by the four fourth roots of unity. The number of bands with a particular eigenvalue at the origin is a topological invariant, because changing it would either require closing the gap or breaking the symmetry. In particular this means that at the origin we already have four topological invariants, one for each eigenvalue. Repeating this procedure for other high-symmetry points of the Brillouin zone as well, yields other topological invariants, but these are not all independent. There are gluing conditions between representations associated to points and ones associated to other subspaces such as lines and planes. In other words, when going from a point with less symmetry to a point with more symmetry, the representations of the bigger stabilizer group have to restrict to the representations of the little group. This becomes especially visible when considering two-dimensional crystal groups with reflection symmetries that cause full circles to be fixed (or even 2-tori in three dimensions). On these circles there can then be special points at which the stabilizer group enhances. 

Implementing these constraints consistently on the full set of topological invariants on each fixed point determines the local invariants. To complete the classification we also have to determine the global invariants. These invariants naturally live on circles or surfaces and at the same time use the global topology of the Brillouin zone in a non-trivial way. They generalize the more well-known invariants to the case where the phase is protected by additional crystal symmetries. 

The global invariants can be visualized in an intuitive way as follows. Let us start with a basic example, the Chern number, which is known to model the plateaus of the IQHE. 
While not really necessary for this case, we will for simplicity assume that the 
Brillouin zone is a two-dimensional sphere rather than a two-torus. With this assumption we are ignoring most of 
the topology of the Brillouin zone, with would correspond to a system which 
is neither protected by time-reversal symmetry nor by any other symmetry. The sphere can then for example
be thought of as a compactified version of the entire two-dimensional space of momenta. We will frequently 
encounter spheres in what follows, which is why we choose this approximation here as well, but we emphasize
once more that for this particular case the result does not depend on whether we take the
Brillouin zone to be a two-sphere or a two-torus.

For the two-sphere (or two-torus), it is well known that there exists an infinite family of non-trivial band structures labelled by the integral of the Berry curvature \cite{Ryu2010}. This integral is an integer and is known as the Chern number. Explicitly, it can be written as 
\be\label{Chern}
C = \frac{1}{2\pi} \int_{S^2} \Tr F  \in \mathbf{Z}
\ee 
with $F$ the Berry curvature two-form constructed out of a $U(N)$ Berry connection $A^{\a\b}_{i} = \braket{\a|\partial_i|\b}$, where $\a = 1, \dots N$, $\ket{\a}$ are the energy eigenstates of a Hamiltonian $H$ and $\partial_i$ derivatives along the directions of the sphere. The curvature is $F = dA + A \wedge A$, which is again $U(N)$-valued. Let us focus on $N = 1$. In that case we have a $U(1)$-connection on the sphere. To see what type of connections correspond to non-trivial Chern numbers and hence to non-trivial topological invariants, we concentrate all the curvature on the north pole. This is possible as long as we do not change the integral. In fact, in the limit we simply have a delta function at the north pole with a coefficient such that the integral in \ref{Chern} is an integer. On the level of the connection we can view this configuration as a vortex around the north pole. In local coordinates $\varphi$ around that point, the connection will simply be
\be
A = d\varphi.
\ee
We can thus conclude that a non-trivial Chern number corresponds to a vortex in the Berry connection. The location of these vortices can be moved (the north pole is not special) but their vorticity is a topological invariant. 

The situation becomes more interesting when we consider topological insulators with time-reversal symmetry, which come with their associated $\Z_2$ invariants. Using the vortex picture sketched above, we can understand this invariant as follows. When there is a time-reversal symmetry present which squares to minus one, the band structure will always consist of an even number of bands. Thus the minimal Berry connection is at least $U(2)$-valued. However, the curvature $F$ will always be zero, since time-reversal symmetry acts as an orientation reversing operation on the base space, i.e the sphere. Nevertheless, this does not mean that there is no other topological invariant. As shown by \cite{DeNittis2015, Ryu2010}, there still exists a $\mathbf{Z}_2$ invariant. If we focus on the $U(2)$ case, the upshot of \cite{DeNittis2015} is that there is still a way to define a Chern number associated to only one of the two energy eigenstates. Its parity then gives the $\mathbf{Z}_2$ invariant. The other energy eigenstate will then carry the opposite Chern number. In local coordinates around e.g. the north pole, the non-trivial connection will then look like
\be
A = \begin{pmatrix}
d\varphi  &  0\\
0 & -d\varphi
\end{pmatrix}
\ee
or any other odd vorticity in each block.
The trivial connection instead has an even vorticity in both blocks. The non-trivial connection is thus one of a vortex-antivortex pair in the Brillouin zone. Notice that their position can be changed, but due to time-reversal symmetry they are always at antipodal points. Of course this is a simplified picture, but it serves as an intuitive and physical interpretation of the invariant. In particular, when other crystal symmetries are added, the vortices need to respect that symmetry too. This greatly constraints their position and in combination with representation theory, their vorticity \cite{Kruthoff2}. 

%
%

\subsection{Outline, summary of results and comparison}

The objective of this paper is to formalise some of these ideas. In particular, in Section \ref{sec:time-reversal} we will introduce some basic terminology in algebraic topology by discussing an example of a crystal with only time-reversal symmetry. The next section, Section \ref{sec:spectralseq} contains the meat of the paper. We discuss in more detail what $K$-theory we want to compute to classify topological insulators with time-reversal and crystal symmetry. To compute these K-theories we describe the construction of an Atiyah-Hirzebruch spectral sequence and compute two examples in detail. Section \ref{sec:generalisations} is devoted to various other examples we computed, for example, we compute, for the first time, the full classification of a two dimensional crystal with time-reversal in class AII and a four fold rotation symmetry. Furthermore, we also determine the twisted representation rings, which are needed in the spectral sequence, in an algorithmic way. In Section \ref{discussion} we mention various subtleties and future directions. Finally, in appendix \ref{appendix} we have gathered various mathematical details on the spectral sequence construction, twists and twisted group algebras.   

Computing twisted equivariant $K$-theory groups using an Atiyah-Hirzebruch spectral sequence is not new. In previous works, \cite{Shiozaki:2018srz, Shiozaki:2018yyj}, an Atiyah-Hirzebruch spectral sequence was also proposed and used to compute the classification for certain symmetry groups and classes. In this work, we fill in certain gaps left open in these works and put the computation of the K-theory groups with an Atiyah-Hirzebruch spectral sequence on a firm mathematical footing. We have gathered most of these details in the appendix.

The K-theory groups we have computed match with known results in the literature, but also agree with a set of heuristic arguments given in \cite{Kruthoff1, Kruthoff2} in the cases where we have explicit results. In particular, for the Altland and Zirnbauer classes AI and AII with an order two symmetry in two dimensions, our results match with those in \cite{PhysRevB.88.075142, PhysRevB.90.165114, PhysRevB.88.125129}. These works extended the analysis by Kitaev in \cite{Kitaev2009} to include additional order two symmetries such as a reflection or two fold rotation symmetry. The basis of this analysis is Clifford algebras, which allow for a straightforward implementation of order two symmetries, but for more complicated symmetries, such a procedure is more difficult. In those cases one has to resort to more sophisticated computational methodes of which we outline one in this paper. 

\section{Time-reversal only} \label{sec:time-reversal}

In this section we shall focus on topological insulators with only unbroken time-reversal symmetry on a two-dimensional lattice without any additional rotation or reflection symmetries. Such topological phases belong to either symmetry class AI or AII \cite{Ryu2010}. In the former case the time-reversal operator $T$ squares to the identity, whereas in the latter case it squares to minus the identity. To classify such topological insulators, we need to know how many topologically distinct insulators there are with this symmetry. As was explained in the introduction, with distinct we mean that upon going from one to the other phase, either the gap closes or the symmetry is broken. For a more formal definition, see \cite{Freed2013}.

The classification is most easily understood by translating the problem to momentum space, where discrete translations cause the momenta to only take values in a two-dimensional Brillouin torus $\mathbf{T}^2$. We visualize this torus as the square $[-\pi,\pi] \times [-\pi,\pi]$ with opposite sides identified. Due to time-reversal symmetry, a non-trivial group acts
on this two-torus which sends $k$ to $-k$, which is intuitively clear as reversing time should reverse the sign of momenta.
Time-reversal symmetry can also easily be shown to be an anti-unitary symmetry. Another example of a possible 
anti-unitary symmetry (which we will not consider in this paper) is particle-hole symmetry, which acts trivially on the
torus. Since we will ignore interactions the momenta $k$ are conserved quantities that can be used to label the states in our Hilbert space. The states with label $k$ are exactly the momentum $k$ Bloch waves. The collection of all these Hilbert spaces form a vector bundle. This vector bundle is the collection of all valence and conduction bands and since we are dealing with insulators here, there is a gap between them. In a (topological) insulator, only the valence bands are physically relevant. For the classification, we hence focus on this finite-dimensional sub-bundle. 

The classification of topological insulators has now been translated into a mathematical question about the classification of vector bundles over the torus. In the absence of time-reversal symmetry, such a classification can be performed using standard (complex) $K$-theory. With time-reversal symmetry, things get a little more exotic, since time-reversal is an antiunitary operator which in particular anticommutes with the imaginary unit $i$. Nevertheless, Atiyah \cite{atiyahKR} generalized $K$-theory to incorporate this symmetry and dubbed it Real $K$-theory. Specifically, for class AI and AII, we need to compute $KR^{-q}(\mathbf{T}^2)$. Here the $\mathbf{T}^2$ is the two-dimensional Brillouin zone and the index $q$ labels the various Altland-Zirnbauer classes \cite{altlandzirnbauer}. In this situation we need $q =0,4$ as they indicate class AI and AII respectively. It is actually not too hard to compute these $K$-theory groups \cite{Kitaev2009, Freed2013}. The result is 
\be
\label{K-theory of 2-torus}
KR^0(\mathbf{T}^2) = \mathbf{Z},\quad KR^{-4}(\mathbf{T}^2) = \mathbf{Z}\oplus \mathbf{Z}_2.
\ee
The conventional computation of these groups uses various basic properties of $KR$-theory, which cannot be generalized to include point group symmetries.
Moreover, this computation is rather unsatisfactory as it gives no insight into what these invariants mean and where they come from. Part of the motivation of this work and of \cite{Kruthoff1,Kruthoff2} is to understand what the physical origin is of these invariants and what computational tool makes this physical origin manifest. In particular, we would like to see how the gluing of representations reveals itself in the computation. Looking ahead, we can interpret the result \eqref{K-theory of 2-torus} as follows. The invariants $\mathbf{Z}$ are local in nature and just give the rank of the bundle, i.e. they represent the number of valence bands present. The more interesting invariant $\mathbf{Z}_2$ is a global two-dimensional strong topological invariant called the Fu-Kane-Mele invariant and is related to topologically protected edge states \cite{FuKaneMele, KaneMele}.


In order to better understand the physical origin, we decompose the Brillouin zone into various parts that are easy for $K$-theory to handle. Within $K$-theory we have the freedom to consider a so-called stable equivalent space instead of the torus. Fortunately, there exists a nice space that is stably equivalent to the torus. This space is a certain wedge sum of one and two-dimensional spheres\footnote{The wedge sum of two spaces is the union of the two spaces but where one point of the first space is identified with one point of the second.}. Moreover, $K$-theory is additive under taking such wedge sums and hence we only have to compute the $K$-theories of spheres, see the end of Section \ref{sec:general method} for a more precise statement. Physically, this means that we are looking at properties of the band structure insensitive to (part of) the discrete translation symmetry. The two-dimensional sphere just represents the (compactified) momentum space of a topological insulator without translation symmetry and the $K$-theory then gives the topological invariants associated with this Brillouin zone. For instance, the Chern number in the IQHE is just the complex $K$-theory of the $2$-sphere and is known not to rely on translational symmetry. After computing the $K$-theories of all such pieces, one simply assembles all these pieces together by taking direct sums. 

In two dimensions, going from the torus to the sphere can be accomplished by identifying the boundary of the square $[-\pi,\pi]\times[-\pi,\pi]$ with a single point. Let us focus on this sphere for the moment.   
After this operation, the time-reversal action of $\mathbf{Z}_2$ on the Brillouin zone torus reduces to an action on the sphere that is still given by the formula $k \mapsto -k$ if we view the sphere as $\mathbf{R}^2 \cup \infty$.
Now suppose we have a Hilbert bundle $\mathcal{H}$ over the sphere with time-reversing operator $T$, i.e. a bundle map $T: \mathcal{H}_k \to \mathcal{H}_{-k}$, where $\mathcal{H}_k$ denote the fibers of the bundle $\mathcal{H}$.
There are two special points at $k = 0$ and $k = \infty$ under the action of time-reversal at which the Bloch states with momentum $k$ are mapped to themselves. 
This gives vector space automorphisms on the corresponding fibers $\mathcal{H}_k$.
In class AI (so $T^2 = 1$), the operator $T$ acts as an effective complex conjugation on the Bloch states of momenta $k = 0$ and $k = \infty$. 
In more mathematical jargon, there are canonical real structures on the vector spaces $\mathcal{H}_0$ and $\mathcal{H}_\infty$.
In class AII, when $T$ squares to $-1$, we instead have canonical quaternionic structures at $k = 0$ and $k = \infty$.
In particular, we deduce that the space of Bloch waves at these special points is even dimensional, which is a manifestation of Kramer's theorem.
However, at a generic point on the sphere, the momenta are not preserved by $T$, so that the state spaces at these points do not admit any extra structure. 

We have now discussed how time-reversal acts on the Brillouin zone once the torus is reduced to a sphere. To include more complicated symmetries later on, it is convenient to view the sphere as being build up out of points, intervals and disks. We have chosen these particular building blocks because they are topologically trivial, i.e. contractible. Such a collection of building blocks is called a CW-complex. The building blocks themselves are called $d$-cells where $d$ is the dimension of the block. When additional symmetries are present, such a CW-complex has to respect the symmetry.
By this is meant that for each cell the symmetry must either fix it completely or map it to a different cell in the decomposition.
In the case of time-reversal symmetry for instance, such a CW-complex is given in Figure \ref{fig:CW structure}.   
In this figure, we also gave the cells an orientation that is preserved by the symmetry, which is visualized by the direction of the arrows on the $1$-cells.
Note that the north and south pole are fixed by the $\mathbf{Z}_2$ action and hence constitute the $0$-cells. 
The $1$-cells are a line from $p_0$ to $p_{\infty}$ and its symmetry-related partner. The same is true for the $2$-cells, which are the two hemispheres.
This yields a practical setting to do the classification using $K$-theory, because we can simply classify the bundles over these $d$-cells and then glue them together consistently. Let us see how this works in more detail. 

To start, we consider the complex and Real K-theory of spheres. It is a known fact that the K-theory of a point in degree $-p$ is equal to the (reduced) K-theory of a $p$-dimensional sphere. The results are given in Table \ref{TwistedRing}. 
Bundles on the $0$-cells, i.e the north and south pole in Figure \ref{fig:CW structure}, are classified by $KR^0(\pt)$. In class AI, we have $KR^0(\pt) = \mathbf{Z}$ for each of the two fixed points. The $\mathbf{Z}$ now assigned to the north and south pole are simply given by the dimension of the fiber at those points. In class AII we get for each fixed point $KR^{-4}(\pt) = \mathbf{Z}$, which is given by the quaternionic dimension of the fiber. On the two intervals there is no real or quaternionic structure. Hence we should assign the (reduced) complex K-theory of the interval, where the boundary points of the interval are identified with each other. This K-theory is equal to the (reduced) complex K-theory of the circle, which is zero. The precise reason for this assignment is addressed in detail in the appendix. 
Finally to the two hemispheres, we assign the (reduced) complex K-theory of a sphere, which is $\mathbf{Z}$. As before, the sphere appears here because we are identifying the boundary of the disc to a point. If our Hilbert space of states is to be preserved by the time-reversal symmetry, the bundle over the two 2-cells should come in pairs that are mapped into each other by the action of time-reversal symmetry. It is thus enough to know the bundle on one such 2-cell and hence under $T$, the two copies of $\mathbf{Z}$ are identified. We thus have $\mathbf{Z}^2$ in zero dimensions ($0$-cells), a $0$ in one dimension ($1$-cells) and $\mathbf{Z}$ in two dimensions ($2$-cells).

To get to a complete classification of topological insulators, we have to make sure that our assignment of bundles to cells is consistent.
This can be done by imposing constraints in successive dimensions.
For dimension zero, this means that when the fibers above the $0$-cells are all extended to the $1$-cells, the result should be consistent.
In our case this means that the state spaces at the points $k = 0$ and $k = \infty$ should have the same dimension, thereby reducing the $\mathbf{Z}^2$ we found before to $\mathbf{Z}$.

This approach is intuitively clear and can easily be generalized to include point group symmetries.
However, as advocated in the beginning, the approach of assigning representations to points is only part of the full classification. To get the other part, the global part, we should check consistency of assignments of bundles (not just representations) to higher-dimensional cells. This becomes a lot more difficult and it is hard to understand for generic crystal symmetries. In the case without time-reversal symmetry, these invariants are most of the time first Chern numbers, but there are exceptions \cite{Shiozaki:2018srz}. 
The invariants that can take any integer value can be understood by using the equivariant Chern character \cite{cherncharacter} or Segal's formula \cite{hirzebruch}, which also has an extension to the twisted case \cite{Adem:2001gv}. However, for crystals invariant under time-reversal symmetry, the invariants are often torsion invariants and take only particular integer values. There is no systematic way of understanding them in the sense that there is no explicit formula for this piece. Instead, when assigning bundles to higher-dimensional cells, we have to check which bundles can be realized as a certain cohomological boundary and quotient out by these.
The result will indeed give the $\mathbf{Z}_2$-invariant of equation \eqref{K-theory of 2-torus}, but it requires some abstract mathematical theory to see this.
Physically, however, there is a heuristic way of understanding these invariants as vortex-anti-vortex pairs in the connection on the bundle, which was presented in the beginning of this section. 

Below we will formalize the heuristic arguments given above and put them on a firm mathematical footing. The example we have seen in this section will be computed again using machinery that allows for a generalization to more complicated crystal symmetries. To illustrate this, we compute the full classification for topological insulators in class AII on a two dimensional lattice with a twofold rotation symmetry.

\section{The spectral sequence and applications}\label{sec:spectralseq}

Now we come to the core of the paper. In the above we gave a heuristic classification of topological insulators with time-reversal symmetry. We will now make this classification precise and generalize to cases with non-trivial crystal symmetry. The strategy of this section will be to introduce all necessary tools. We will then reconsider the example without any crystal symmetry but with time-reversal symmetry. Whenever appropriate, we will mention the physical motivation and interpretation for these tools along the way.  

%

Let us consider topological insulators in $d$ dimensions in class A, AI or AII, possibly with a point group symmetry. We denote the full classical symmetry group of the Brillouin zone by $G$. If present, $G$ therefore contains time-reversal symmetries and point group symmetries but no translational symmetries. These are taken into account by the topology of our Brillouin zone torus. Let us denote by $\mathcal{G}$ the space group and $G$ its (magnetic) point group that does not contain time-reversal symmetry, then the space group $\mathcal{G}$ is a group extension
\be
1 \to \Z^d \to \mathcal{G} \to G \to 1,
\ee 
where $\Z^d$ represents the discrete lattice translations in $d$ spatial dimensions. When this extension is split, the space group is called symmorphic and non-symmorphic otherwise. We will focus on the former from now on and comment on the non-symmorphic case in the discussion. We will assume that there are no other symmetries, such as gauge symmetries with which the time-reversal operator could mix. 

In order to classify topological phases in the sense of Freed and Moore \cite{Freed2013}, we have to compute a joint generalization of Real and equivariant $K$-theory. In particular, we want to take two additional things into account. First of all, we want to keep track of which elements in $G$ act antiunitarily or not. For this we will use a map $\phi$ which sends an element of $G$ to $+1$ if it is unitary and to $-1$ if it is antiunitary. Moreover, we want to know how elements of $G$ acting on the Brillouin zone lift to elements acting on the fiber. This is most easily accounted for by a twist $\t$, a suitable group two-cocycle.
This twist encodes the action of the symmetries on the quantum Hilbert space.
For example, it prescribes whether $T^2 = \pm 1$.
But it also provides the signs coming from taking the spin of particle into account. For example, an $n$-fold rotation operator $R$ for spin$-\tfrac{1}{2}$ particles satisfies $R^n = -1$. This minus sign is also encoded in $\t$. 

Let us for a moment describe the situation in more precise abstract mathematical terms.
Assume we have the following data:
\begin{enumerate}[(i)]
\item a finite group $G$ acting on a space $X$ (in our case $X=\mathbf{T}^d$, the Brillouin zone);
\item a homomorphism $\phi:G\to \mathbf{Z}_2$;
\item a group $2$-cocycle $\tau\in Z^2(G,U(1)_\phi)$ with values in the circle group $U(1)$ with $G$-action $g \cdot e^{i\theta} = e^{i\phi(g) \theta}$.
\end{enumerate}
Such a cocycle $\t$ is a special case of the more general twists defined by Freed and Moore \cite{Freed2013}, called $\phi$-twisted central extensions.
Using such data, Freed and Moore \cite{Freed2013} defined a version of twisted equivariant $K$-theory denoted by
\be
^\phi K^{\t}_G (X),
\ee
which was further studied in \cite{gomi}. 
It was also argued that this $K$-theory group classifies free fermion topological insulators protected by the quantum symmetry defined by $G,\phi$ and $\t$.

To connect with  more common language used in the physics literature, we describe the $G,\phi$ and $\t$ that occur in the classification of crystalline topological insulators.
Firstly, a class A topological insulator with point group $G_0$ simply has $G = G_0$ and $\phi$ and $\t$ are both trivial.
For class AI, $G$ will instead be the magnetic point group, i.e. it will contain both point symmetries and time-reversal symmetry.
We will only consider magnetic point groups of the form $G = G_0 \times \Z_2^T$, with $\Z_2^T$ the action of time-reversal symmetry on the Brillouin zone, even though the mathematical machinery developed here can handle more general point groups as well. For instance, one could also consider cases in which the time-reversal operator is a combination of the usual time-reversal operator with a lattice translation or point group symmetry in $G_0$. 
For $G = G_0 \times \Z_2^T$ in class AI, $\phi: G \to \Z_2$ will simply be projection onto the second factor and $\t$ will be trivial.
Finally, for class AII, we again take $G = G_0 \times \Z_2^T$ to be the magnetic point group and $\phi$ the same projection, but now we pick $\t$ in such a way that the twisted group action represents the desired action on the quantum Hilbert space.
In particular, we pick $\t$ so that time-reversal squares to $-1$, reflections square to $-1$ and rotation by $2 \pi$ equals $-1$.
To assure a consistent choice, a precise construction of $\t$ for a given point group $G_0$ is given at the end of appendix \ref{sec:Freed Moore twists}.

It is shown in \cite[Thm 3.11]{gomi} that the groups $^\phi K^{\t}_G (X)$ satisfy certain equivariant versions of the homotopy, excision, additivity and exactness axioms of Eilenberg and Steenrod.
The fact that our twisting class is defined by a group cocycle implies that these axioms are exactly the axioms for an equivariant cohomology theory on the category of $G$-spaces as defined in Bredon \cite[\S I.2]{bredonequivariant}. This is what makes the following computations mathematically sound; as explained in \cite[\S IV.4]{bredonequivariant} the axioms guarantee the existence of the Atiyah--Hirzebruch spectral sequence. Moreover, the orbifold point of view advocated in \cite{gomi} allows us to change the group $G$ and the space $X$ as long as the quotient space remains the same and we keep the same stabilizer.
This is useful in some computations, see the end of Section \ref{sec:rotation}.
For more details on how the $K$-theory we use is defined, see appendix \ref{sec:Freed Moore twists}.

We are therefore left with the task to compute the $K$-theory $^\phi K^{\t}_G (\mathbf{T}^d)$ of the Brillouin zone dressed with $\phi$ and $\t$. The technique to compute these groups goes along the lines that we have discussed in the previous section. 
We first decompose the Brillouin zone into cells and view them as a CW-complex. Non-trivial symmetries have to leave these complexes invariant. Such complexes are equivariant $G$-CW complexes, which is nothing more than an upgraded version of the unit cell in momentum space. After having found this $G$-CW complex, we use an Atiyah-Hirzebruch spectral sequence to compute $^\phi K^{\t}_G (S^i)$, which are assembled to give $^\phi K^{\t}_G (\mathbf{T}^d)$. Let us now formalize this computational method. 

\subsection{A general method: the Atiyah-Hirzebruch spectral sequence} \label{sec:general method}

The spectral sequence for the computation of the twisted equivariant $K$-theory of a space $X$ is constructed by using a decomposition as a $G$-CW complex $X^0 \subseteq \cdots \subseteq X^d = X$, where the superscript on the cells indicates the dimension of the subspace. For the applications considered in this paper, $X$ is either a torus $\mathbf{T}^d$ or a sphere $S^d$ (we will remark on how to reduce the computation of the $K$-theory of the torus to the $K$-theory of a sphere at the end of this section). 
A spectral sequence is a successive approximation method converging to the desired answer in a number of steps. For us these steps will always be finite and in fact, most of the time only two steps are necessary. These steps are usually referred to as pages. The first page of the spectral sequence, just as in the last example, is given by equivariant assignments $f$ of the $K$-theory of spheres to the cells of $X$. For the $0$-cells $X^0$, this means that we assign to each $k \in X^0$ a twisted representation of the stabilizer group $H_k$ of that point. These representations, which are twisted using $\t$ and the map $\phi$, are conveniently packaged in the twisted representation ring $^\phi R^\tau(H_k)$. These objects are actually not rings, but since they are equal to the usual representation ring of $H_k$ in case $\phi$ and $\t$ are trivial, we will keep on referring to them as twisted representation rings. Details about twists and twisted representations can be found in appendix \ref{sec:Freed Moore twists}. So $f$ maps $k$ to an element of the twisted representation ring $^\phi R^\tau(H_k)$ of the corresponding stabilizer group $H_k$. By equivariance is meant that $f$ preserves the symmetry in the following sense: $f$ is required to map $gk$ to the resulting conjugate representation in $^\phi R^\tau(gH_kg^{-1}) = {}^\phi R^\tau(H_{gk})$.
More generally, we equivariantly assign higher representation rings $^\phi R^{\tau - q}(H_\sigma)$ (i.e. the higher degree twisted equivariant $K$-theory of a point, see appendix \ref{sec: twisted reprings} for details) to $p$-cells $\sigma$. These classify twisted $H_\sigma$-equivariant bundles over $q$-spheres, instead of over just a point. The grid of such assignments of representation rings for each $p$ and $q$ form the first page of the spectral sequence and is denoted by $E^{p,-q}_1$.
Those assignments can be shown to be equivalent to Bredon $p$-cochains with values in the coefficient functor $^\phi \mathcal{R}^{\tau - q}_G$. In appendix \ref{sec:spectral sequence} we define these coefficient functors and present a derivation of this result. Intuitively, the functor $^\phi \mathcal{R}^{\tau - q}_G$ keeps track of both the (higher) representations at fixed loci and how they restrict to each other. For Bredon $p$-cochains, this functor will pick out the stabilizer group of the $p$-cells and assign degree $-q$ twisted representation rings to the $p$-cells.  
It should be noted here that the action of group elements on the higher representation rings can be tedious to determine explicitly in certain examples, so that the equivariance of $f$ can result in nontrivial results. One example of this is given in Section \ref{sec:revisited}.

To go to the next page of the spectral sequence, we have to take the cohomology of the first page with respect with the first differential, which in our case is known as the Bredon differential. In fact, the first differential maps $E^{p,-q}_1$ to $E^{p+1,-q}_1$ and is given by the differential of Bredon cohomology, which is
\be\label{differential}
(d f)(\sigma) = \sum_{\m \in C^p(X)} [\m : \sigma] f(\m)|_{G_\sigma},
\ee
with $C^p(X)$ denoting the set of $p$-cells of $X$ and $f$ a Bredon $p$-cochain. 
Here $f(\m)|_{G_\sigma}$ means that we take the higher twisted representation of $G_\m$ that $f$ assigns to $\m$ and restrict it to a representation of $G_\sigma$.
The notation $[\m : \sigma]$ stands for an integer factor that tells us in which way $\m$ intersects the boundary of the $p+1$-cell $\sigma$.
In general the behavior and computation of this number can be quite complicated, but if our CW-complex is sufficiently nice, this number is usually just a sign depending on a fixed orientation.
For example, if we have a line (1-cell) $\ell$ oriented from the endpoint $p_0$ to the other endpoint $p_1$, i.e.  $\partial \ell = p_1 - p_0$, then we simply have
\be
[\ell:p_1] = 1, \quad [\ell:p_0] = -1
\ee
and of course $[\ell:\pt] = 0$ if $\pt$ is not an endpoint of $\ell$.
If instead $\sigma = A$ is a $2$-cell that lies in a disk surrounded by a couple of intervals $\ell_1, \dots, \ell_k$, then $[A:\ell_i] = \pm 1$ depending on whether the orientations of the line $\ell_i$ coincide with the orientation of $A$.
In more general situations, where there is nontrivial gluing present, it can be computed as the degree of a certain map between spheres. This map is exactly the same as for the cellular boundary map in ordinary cellular homology, which can be found for example in Hatcher's book \cite{hatchertopology}.

The second page is the cohomology of the first page with respect to the differential $d : E_1^{p,-q} \to E_1^{p+1,-q}$ given in \eqref{differential}.
Mathematically speaking the second page entry $(p,-q)$ therefore equals the degree $p$ Bredon equivariant cohomology of $X$ with coefficient functor $^\phi \mathcal{R}^{\tau - q}_G$.
For the third and higher order pages, we need to know the higher differentials, which are much more abstractly defined and no explicit form is known. Therefore, until more is known about this it is not possible to fully classify topological phases for general point groups using this method. It is however often the case in practice that we can arrive at a definite answer without knowing explicit expressions for the higher differentials. At least it is known that the $r$th differential is of bidegree $(r,1-r)$, so $d_r : E_r^{p,-q} \to E_r^{p+r,-q+1-r}$.
Therefore, for $d$-dimensional spaces, the $r$th differential $d_r$ is zero for all $r>d$.  
For more details on the construction of the spectral sequence and explicit definitions, see appendix \ref{sec:spectral sequence}.

But how do we construct the twisted equivariant $K$-theory of $X$ from the data of the spectral sequence?
After taking the cohomology with respect to the $d$th differential, we arrive at the final page, $E^{p,-q}_\infty$.
We can construct the $K$-theory by extensions out of $E^{p,-p}_\infty$, so by equating $p$ and $q$.
In two dimensions, this means that there exist exact sequences
\begin{align} \label{eq: exact sequences1}
    0 \to E^{2,-2}_\infty \to F \to E^{1,-1}_\infty \to 0,
    \\\label{eq: exact sequences2}
    0 \to F \to {}^\phi K^{\t}_G (X) \to E^{0,0}_\infty \to 0,
\end{align}
see the final paragraph of appendix \ref{sec:spectral sequence} for the details.
Unfortunately, these sequences do not split in general. Therefore the $K$-theory is not always fully determined by the spectral sequence (unless of course we would explicitly determine the maps in these sequences, which is a tedious exercise).
We will call this the problem of non-unique extensions, which unfortunately is intrinsic to our approach.
An example of this phenomenon will be addressed in Section \ref{sec:rotation}.

Now that the spectral sequence is contained in our toolbox, we will explain how to reduce the computation of equivariant $K$-theory of the Brillouin torus $\mathbf{T}^d$ to the computation of the $K$-theory of spheres.
For this we use an equivariant stable homotopy equivalence that generalizes \cite[Thm 11.8]{Freed2013}. This equivalence adresses the decomposition of the Brillouin torus in terms of spheres. 
Indeed, if the action of $G$ on $\mathbf{T}^d = S^1 \times \dots \times S^1$ can be realized as the restriction of an action of $H^d \rtimes S_d$, where $H$ acts on $S^1$ and $S_d$ permutes the copies of $S^1$, then $\mathbf{T}^d$ is equivariantly stably homotopy equivalent to a wedge of spheres.
More explicitly, this means in two dimensions that there is an isomorphism
\be\label{eq: equivariantsplittingb}
\leftidx{^\phi}{K}{^{\tau}}_G(\mathbf{T}^2)
\cong  \leftidx{^\phi}{K}{^{\tau}_G}(S^2) \oplus \leftidx{^\phi}{\widetilde{K}}{^{\tau}_G}(S^1 \vee S^1).
\ee
Here the tilde indicates the reduced $K$-theory and $S^1 \vee S^1$ is a space that looks like the figure $8$, which is nothing but the boundary of the Brillouin zone torus $\mathbf{T}^2$ seen as a square $[-\pi,\pi] \times [-\pi,\pi]$ with opposite sides identified.
Note that the symmetry $G$ could potentially interchange the two $S^1$'s of the figure eight, for example in case there is a fourfold rotation symmetry. If there is no group element permuting the two copies of the circle, the $K$-theory decomposes further as
\be \label{eq: equivariantsplitting}
\leftidx{^\phi}{K}{^{\tau}}_G(\mathbf{T}^2)
\cong \leftidx{^\phi}{K}{^{\tau}_G}(S^2) \oplus \leftidx{^\phi}{\widetilde{K}}{^{\tau}_G}(S^1) \oplus 
\leftidx{^\phi}{\widetilde{K}}{^{\tau}_G}(S^1),
\ee 
where we used that
\be
\leftidx{^\phi}{\widetilde{K}}{^{\tau}_G}(S^1 \vee S^1) = \leftidx{^\phi}{\widetilde{K}}{^{\tau}_G}(S^1) \oplus \leftidx{^\phi}{\widetilde{K}}{^{\tau}_G}(S^1).
\ee
A similar isomorphism as in \eqref{eq: equivariantsplitting} exists in three dimensions under the given assumptions. The relation between reduced and unreduced $K$-theory $\leftidx{^\phi}K{^{\tau}_G}(X)$ is
\be
\leftidx{^\phi}K{^{\tau}_G}(X) = \leftidx{^\phi}{K}{^{\tau}}_G(\pt) \oplus \leftidx{^\phi}{\widetilde{K}}{^{\tau}_G}(X).
\ee
When using the equivariant splittings \eqref{eq: equivariantsplittingb} and \eqref{eq: equivariantsplitting}, we can thus compute the unreduced $K$-theory and then strip of the $\leftidx{^\phi}{K}{^{\tau}}_G(\pt)$-part to obtain the reduced $K$-theory.
Note that the assumption that the action of $G$ comes from some action of $H^d \rtimes S_d$ does not always hold, so that we cannot always use \eqref{eq: equivariantsplittingb}.
If for example three-fold rotations are present, we seem to be bound to applying the Atiyah-Hirzebruch spectral sequence to the Brillouin zone torus directly.

Since the $K$-theory of a one-dimensional space $X$ is easy to compute, the isomorphism \eqref{eq: equivariantsplitting} effectively reduces computations of the $K$-theory of a two-dimensional torus to a two-dimensional sphere.
Indeed, for one-dimensional spaces all higher differentials vanish and $E^{2,-2}_\infty = 0$, so that the exact sequences \eqref{eq: exact sequences1} and \eqref{eq: exact sequences2} reduce to a single exact sequence.
Because the twisted representation ring ${}^\phi R^\tau(G)$ is torsion free (for $q = 0$), so is $H^0(X,^\phi\mathcal{R}^{\tau}_G)$. 
Hence the resulting sequence splits, giving us
\be \label{eq: one-dimensional K-theory}
{}^\phi K^{\t}_G (X) \cong
H^0(X,^\phi\mathcal{R}^{\tau}_G) \oplus H^1(X,^\phi\mathcal{R}^{\tau-1}_G).
\ee
Despite the absence of torsion in the first term, the second term can give rise to torsion of which we will see examples below. The torsion in $H^1$ was anticipated before in \cite{PhysRevB.94.165164} for systems in class AII with a reflection symmetry in one dimension. 

\begin{figure}[t]
    \centering
\begin{tikzpicture}[scale=1.5]
\begin{scope}[decoration={markings,mark=at position 0.5 with {\arrow{<}}}]
\draw[dashed,postaction = {decorate}] (0,1) arc (90:270:0.5cm and 1cm);
\draw[postaction = {decorate}] (0,1) arc (90:-90:0.5cm and 1cm);
\end{scope}
\draw (0,0) circle (1cm);
\draw [fill] (0,-1) circle [radius = 0.05];
\draw [fill] (0,1) circle [radius = 0.05];
\node[below] at (0,-1) {$p_0$};
\node[above] at (0,1) {$p_{\infty}$};
\node at (0.3,0) {$\ell$};
\node at (-0.25,0.2) {$T\ell$};

\path [ball color=blue!40!white,opacity=0.20]  (0,1) arc [radius = 1, start angle = 90, end angle = 270] (0,1) arc(90:-90:0.5 and 1);
\path [ball color=red!40!white,opacity=0.20] (0,1) arc(90:270:0.5 and 1) (0,-1) arc [radius = 1, start angle = -90, end angle = 90];

\draw[->] (-2,0) to[bend left] (-0.8,0); 
\node[left] at (-2,0) {$TA$};
\draw[->] (2,0) to[bend right] (0.8,0);
\node[right] at (2,0) {$A$};
\end{tikzpicture}
\caption{A $\mathbf{Z}_2$-CW structure of the $\mathbf{Z}_2$-space $S^2$ that is the one-point compactification of the two-dimensional representation of $\mathbf{Z}_2$ given by $k \mapsto -k$. We have denoted this action by $T$.}
    \label{fig:CW structure}
\end{figure}
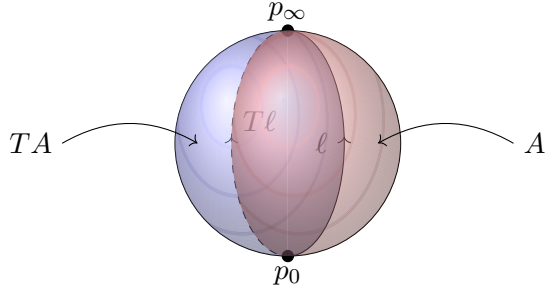

\subsection{Time-reversal only: revisited}
\label{sec:revisited}

Let us now illustrate how the Atiyah-Hirzebruch spectral sequence formalizes the intuitive approach of the last section. So again we will take $X = S^2$ with the $\mathbf{Z}_2$-action $k\mapsto -k$ with the $\mathbf{Z}_2$-CW-structure as given in Figure \ref{fig:CW structure}. 
We will consider the classes AI ($T^2 = 1$) and AII ($T^2 = -1$) simultaneously and note the distinctions along the way.
Mathematically, we distinguish between the two classes by picking the twist $\tau = \tau_0$ to be trivial in class AI and $\tau = \tau_1$ nontrivial in class AII.
The higher twisted representation rings of $\mathbf{Z}_2$ and the trivial subgroup $1 \subseteq \Z_2$ are given in Table \ref{TwistedRing}.
Note that the stabilizers of the 0-cells are both $\mathbf{Z}_2$, while for the other cells the stabilizer is trivial.

We will start by computing all Bredon cohomology groups that are necessary for obtaining the $K$-theory group from the spectral sequence.
These cohomology groups are the ones that correspond to second page entries of the spectral sequence which could possibly influence the three desired entries $E^{0,0}_\infty,E^{1,-1}_\infty$ and $E^{2,-2}_\infty$ of the final page occurring in the exact sequences of equation \eqref{eq: exact sequences1} and \eqref{eq: exact sequences2}.
Because the second differential (which is of bidegree $(2,1)$) is the only possibly nonzero higher differential, we have $E_3 = E_\infty$.
Therefore we have to compute $H^p_G(S^2, {}^{\phi}\mathcal{R}_G^{\t-q})$ for $(p,q)$ equal to $(0,0), (0,1),(1,1),(2,1)$ and $(2,2)$.

\begin{table}
\centering
\begin{tabular}{c|c|c|c|c}
& ${}^\phi{R}{^{\tau_0- q}}(\Z_2)$ 
& ${}^\phi{R}{^{\tau_1- q}}(\Z_2)$ 
& ${}^\phi{R}{^{\tau_0- q}}(1)$
& ${}^\phi{R}{^{\tau_1- q}}(1)$
\\ 
& $=$
& $=$
& $=$
& $=$
\\
&$KR^{-q}(\pt)$
&$KR^{-q-4}(\pt)$
&$K^{-q}(\pt)$
&$K^{-q}(\pt)$
\\
\hline
$q = 0$ & $\Z$  & $\Z$ & $\Z$  & $\Z$
\\
\hline
$q = 1$ & $\Z_2$& $0$ & $0$& $0$
\\
\hline
$q = 2$ & $\Z_2$ & $0$ & $\Z$ & $\Z$
\\
\hline
$q = 3$ & $0$& $0$& $0$ & $0$
\\
\hline
$q = 4$  & $\Z$ & $\Z$ & $\Z$ & $\Z$
\\
\hline
$q = 5$ & $0$& $\Z_2$ & $0$& $0$
\\
\hline
$q = 6$ & $0$ & $\Z_2$& $\Z$ & $\Z$
\\
\hline
$q = 7$ & $0$& $0$ & $0$ & $0$
\\
\hline
$q = 8$& $\Z$ & $\Z$ & $\Z$ & $\Z$
\\
\hline
\vdots & &&&
\end{tabular}
\caption{\label{TwistedRing} The twisted representation rings in the case of $G_{\s} = \mathbf{Z}_2$ and $G_{\s} = {1}$. They are $8$-periodic in the degree.}
\end{table}

First we have to find all necessary Bredon equivariant cochains, as they constitute the first page $E^{p,-q}_1$. 
 We start with $p = q = 0$. So we consider the equivariant $0$-cochains with values in $^\phi \mathcal{R}^{\tau}_G$, which here are the equivariant maps from the set $\{p_0,p_\infty\}$ to ${}^\phi R^{\tau}(\mathbf{Z}_2) = \mathbf{Z}$ for both twists.
 Because the $0$-cells are completely fixed by the group, all $0$-cochains are equivariant.
 Therefore the equivariant $0$-cochains are spanned by two basis elements $\pi_0$ and $\pi_\infty$ over $\mathbf{Z}$:
 \be
 C^0_{\mathbf{Z}_2}(S^2, {}^{\phi}\mathcal{R}_G^\t) = \braket{\pi_0,\pi_{\infty}}_{\mathbf{Z}} = \mathbf{Z}^2.
 \ee
 Here $\pi_0$ maps $p_0$ to $1 \in {}^\phi R^{\tau}(\mathbf{Z}_2)$ and $p_{\infty}$ to $0 \in {}^\phi R^{\tau}(\mathbf{Z}_2)$. For $\pi_{\infty}$ the roles of $p_0$ and $p_{\infty}$ are interchanged. In more basic terms: $\pi_0$ assigns a state space of dimension one to $p_0$ and a zero space to $p_\infty$, while $\pi_\infty$ assigns a zero space to $p_0$ and a one-dimensional space to $p_\infty$.
 
 Going up to $p = 1, q = 0$, there is only one equivariant $1$-cochain, so that
 \be
 C^1_{\mathbf{Z}_2}(S^2, {}^{\phi}\mathcal{R}_G^\t) = \braket{\l}_{\mathbf{Z}} = \mathbf{Z}.
 \ee
 Indeed, from Table \ref{TwistedRing} it is clear the this cochain is an equivariant map from $\{\ell, T\ell\}$ to $\mathbf{Z}$.
 By equivariance, it is uniquely specified by specifying its value on $\ell$, which we take to be $1$ for $\l$. In case the reader is interested in the actual value of $\l(T\ell)$, simply note that the action of $T$ on the representation ring is just complex conjugation.
In case $T$ acts on the representation ring of a nontrivial group this will result in the complex conjugation of nontrivial representations, which are in general not isomorphic to the original representation.
However, a complex vector space is noncanonically isomorphic to its complex conjugation since it has the same dimension.
 Hence the automorphism that $T$ induces on $^\phi R^\tau(1)$ is simply the identity, thus $\l(T\ell) = \l(\ell) = 1$. Along the way we will see that this automorphism is not always trivial and acts with minus the identity on the higher representation ring of degree $-2$. 
 This is the heart of the matter, since it is the aspect that creates torsion in this example.
 
 For $q = 1$, the situation simplifies, since the degree $-1$ representation ring of the trivial group equals zero.
 Hence there are no equivariant $1$-cochains or $2$-cochains with values in $^\phi \mathcal{R}^{\tau - 1}_G$.
 The degree $-1$ representation ring of $\mathbf{Z}_2$ depends on whether the twist $\tau$ is taken trivial (class AI) or nontrivial (class AII).
 For $\tau$ trivial it equals $\mathbf{Z}_2$ and for $\tau$ nontrivial it equals $0$.
 In class AII, we therefore also have no equivariant $0$-cochains with values in $^\phi \mathcal{R}^{\tau - 1}_G$.
 In class AI instead, the equivariant $0$-cochains are spanned by $\pi_0$ and $\pi_\infty$, just as for $q = 0$.
 However, this time they are a basis over $\mathbf{Z}_2$:
 \be
 C^0_{\mathbf{Z}_2}(S^2, {}^{\phi}\mathcal{R}_G^{\t-1}) = \braket{\pi_0,\pi_{\infty}}_{\mathbf{Z}_2} = \mathbf{Z}_2^2.
 \ee
Here $\pi_0$ maps $p_0$ to the nontrivial element of $\mathbf{Z}_2$ and $p_\infty$ to the trivial element, while for $\pi_\infty$ it is the other way around.

Finally, for $q = 2$ there are some subtleties. The degree $-2$ representation ring of the trivial group is equal to $\mathbf{Z}$.
Analogously to $q = 0$, we get that the $1$-cochains are spanned over $\mathbf{Z}$ by a single element $\l$ with $\l(\ell) = 1$. 
Similarly, the $2$-cochains are spanned by a single element $\alpha$ with $\alpha(A) = 1$.
However, unlike for $q = 0$, we have that 
\be\label{nontrivialTaction}
\alpha(TA) = T\alpha(A) = -\alpha(A) = - 1.
\ee
This is because the action of $T$ on the degree $-2$ representation ring is $-1$, as can be shown by an explicit analysis using Clifford algebras, using the explicit definitions in appendix A.2. 
We can conclude that the relevant part of the first page of the spectral sequence for class AI and class AII respectively is given in the following table:

\vspace{3mm}
\begin{center}
\begin{tabular}{c|c|c|c}
& $p = 0$ & $p = 1$ & $p = 2$ \\
\hline
$q = 0$&$C^0_G(S^2,\leftidx{^\phi}{\mathcal{R}}{_G^{\tau_0}}) = \Z^2$    &$C^1_G(S^2,\leftidx{^\phi}{\mathcal{R}}{_G^{\tau_0}}) = \Z$ & \\
\hline
$q = 1$  &$C^0_G(S^2,\leftidx{^\phi}{\mathcal{R}}{_G^{\tau_0 - 1}}) = \Z_2^2$   &$C^1_G(S^2,\leftidx{^\phi}{\mathcal{R}}{_G^{\tau_0 - 1}}) = 0$ & 
  $C^2_G(S^2,\leftidx{^\phi}{\mathcal{R}}{_G^{\tau_0 - 1}}) = 0$ \\
  \hline
$q = 2$     & & $C^1_G(S^2,\leftidx{^\phi}{\mathcal{R}}{_G^{\tau_0 - 2}}) = \Z$
&$C^2_G(S^2,\leftidx{^\phi}{\mathcal{R}}{_G^{\tau_0 - 2}}) = \Z$
\end{tabular}
\vspace{3mm}
\end{center}
\begin{center}
\vspace{3mm}
\begin{tabular}{c|c|c|c}
& $p = 0$ & $p = 1$ & $p = 2$ \\
\hline
$q = 0$&$C^0_G(S^2,\leftidx{^\phi}{\mathcal{R}}{_G^{\tau_1}}) = \Z^2$    &$C^1_G(S^2,\leftidx{^\phi}{\mathcal{R}}{_G^{\tau_1}}) = \Z$ & \\
\hline
$q = 1$  &$C^0_G(S^2,\leftidx{^\phi}{\mathcal{R}}{_G^{\tau_1 - 1}}) = 0$   &$C^1_G(S^2,\leftidx{^\phi}{\mathcal{R}}{_G^{\tau_1 - 1}}) = 0$ & 
  $C^2_G(S^2,\leftidx{^\phi}{\mathcal{R}}{_G^{\tau_1 - 1}}) = 0$ \\
  \hline
$q = 2$     & & $C^1_G(S^2,\leftidx{^\phi}{\mathcal{R}}{_G^{\tau_1 - 2}}) = \Z$
&$C^2_G(S^2,\leftidx{^\phi}{\mathcal{R}}{_G^{\tau_1 - 2}}) = \Z$
\end{tabular}
\end{center}
\vspace{3mm}

With the information we have gathered now, we can construct the second page, consisting of Bredon cohomologies. Let us start off by computing $H^0_G(S^2, {}^{\phi}\mathcal{R}_G^\t)$. For this we need to compute the kernel of the Bredon differential
\be
d: \Z^2 = C^0_G(S^2,{}^\phi \mathcal{R}_G^{\tau}) \to C^1_G(S^2,{}^\phi \mathcal{R}_G^{\tau}) = \Z.
\ee
On the $0$-cochain $\pi_\infty$ it acts as
\be
    d\pi_\infty(\ell) = \pi_\infty(\partial \ell)|_1
    = \pi_\infty(p_\infty - p_0)|_1
    =  \pi_\infty(p_\infty)|_1 - \pi_\infty(p_0)|_1
    = \pi_\infty(p_\infty)|_1,
\ee
    where the symbol $|_1$ denotes the restriction of the representation to the trivial group.
    In class AI, this restriction maps complex vector spaces with a real structure $T$ to their underlying complex vector space.
    Since all complex vector spaces admit a real structure, this implies that the restriction map $^\phi R^{\tau}(\mathbf{Z}_2) \to {}^\phi R^\tau(1)$ is the identity.  
    In class AII, where $T^2 = -1$, the restriction is multiplication by two, because only complex vector spaces of even dimension admit a quaternionic structure. Hence we get
    \be
    d \pi_\infty = 
    \begin{cases}
    \lambda & \text{ if } \tau = 
    \tau_0,
    \\
    2 \lambda & \text{ if } \tau = 
    \tau_1.
    \end{cases}
    \ee
    Using the orientation we analogously get that $d \pi_0 = - d \pi_\infty$.
    In both class AI and AII, we see that the degree zero cohomology equals
    \be
    \ker d = H^0_G(S^2, {}^\phi \mathcal{R}_G^{\tau}) \cong \Z.
    \ee
    In general, this cohomology group contains all local topological invariants.
    More precisely, the zeroth degree cohomology group is actually a mathematical formalization of the heuristic method of consistently assigning representations to point sketched in Section \ref{sec:time-reversal} and used extensively in \cite{Kruthoff1} and \cite{Kruthoff2}.

The next row of the second page is easily deduced from the first page. In class AII, all cochains vanish and therefore so do the cohomology groups. In class AI nontrivial $0$-cochains exist, but not in higher degrees. Therefore the differential is necessarily zero and the second page equals the first page.

The final relevant cohomology group is $H^2_G(S^2, {}^{\phi}\mathcal{R}_G^{\t-2})$. In this case, $T$ induced a non-trivial automorphism on $^\phi R^{\tau-2}(1)$ given by $-1$, so that
\be\label{don1}
d\l(A) = \l(\partial A) = \l(\ell) - T\l(\ell) = 2 \l(\ell) = 2,
\ee
hence $d\l = 2 \a$ for both $\t = \t_0$ and $\t_1$, so that $\im d = 2\Z$. The kernel of $d$ acting on $2$-cochains is $\Z$, since we are in top degree. Thus $H^2_G(S^2, {}^{\phi}\mathcal{R}_G^{\t-2}) = \mathbf{Z}_2$. 

Summarizing the results by filling in the second page of the spectral sequence, we get the following tables for $T^2 = 1$ and $T^2 = -1$ respectively:

\begin{center}
\vspace{3mm}
\begin{tabular}{c|c|c|c}
& $p = 0$ & $p = 1$ & $p = 2$ \\
\hline
$q = 0$&$H^0_G(S^2,\leftidx{^\phi}{\mathcal{R}}{_G^{\tau_0}}) = \Z$    & & \\
\hline
$q = 1$  &$H^0_G(S^2,\leftidx{^\phi}{\mathcal{R}}{_G^{\tau_0 - 1}}) = \Z_2^2$   &$H^1_G(S^2,\leftidx{^\phi}{\mathcal{R}}{_G^{\tau_0 - 1}}) = 0$ & 
  $H^2_G(S^2,\leftidx{^\phi}{\mathcal{R}}{_G^{\tau_0 - 1}}) = 0$ \\
  \hline
$q = 2$     & & &$H^2_G(S^2,\leftidx{^\phi}{\mathcal{R}}{_G^{\tau_0 - 2}}) = \Z_2$
\end{tabular}
\vspace{3mm}

\vspace{3mm}
\begin{tabular}{c|c|c|c}
& $p = 0$ & $p = 1$ & $p = 2$ \\
\hline
$q = 0$&$H^0_G(S^2,\leftidx{^\phi}{\mathcal{R}}{_G^{\tau_1}}) = \Z$    & & \\
\hline
$q = 1$  &$H^0_G(S^2,\leftidx{^\phi}{\mathcal{R}}{_G^{\tau_1 - 1}}) = 0$   &$H^1_G(S^2,\leftidx{^\phi}{\mathcal{R}}{_G^{\tau_1 - 1}}) = 0$ & 
  $H^2_G(S^2,\leftidx{^\phi}{\mathcal{R}}{_G^{\tau_1 - 1}}) = 0$ \\
  \hline
$q = 2$     & & &$H^2_G(S^2,\leftidx{^\phi}{\mathcal{R}}{_G^{\tau_1 - 2}}) = \Z_2$
\end{tabular}
\vspace{3mm}
\end{center}

When $T^2 = -1$, we immediately see that all higher differentials vanish. The spectral sequence thus collapses at $E_2$ and the exact sequences \eqref{eq: exact sequences1} and \eqref{eq: exact sequences2} reduce to the single exact sequence
\be
0 \to \Z_2 \to KR^{-4}(S^2) \to \Z \to 0.
\ee
Since $\Z$ is a free group, the sequence splits.
This gives us $KR^{-4}(S^2) = \Z \oplus \Z_2$. 
Moreover, using the spectral sequence it can easily be shown that $\widetilde{KR}^{-4}(S^1) = 0$ (for an example of a computation of the $K$-theory of a one-dimensional space using the spectral sequence, see the next section).
By the equivariant splitting \eqref{eq: equivariantsplitting}, the $K$-theory of the torus is thus 
\be
KR^{-4}(\mathbf{T}^2) = \Z \oplus \Z_2,
\ee
which confirms the result using a different approach, see equation \eqref{K-theory of 2-torus}.
It is worth noting that the torsion invariant $\Z_2$ managed to appear because of the nontrivial action of $T$ induced by complex conjugation and not by the torsion in the $KR$-theory of a point as it does when computing $KR^{-4}(S^2)$ using the methods of for example Freed and Moore \cite{Freed2013}.

For $T^2 = 1$ another lesson is to be learned from this example.
Namely, note that as long as we do not know any expression for the second differential $d_2: \Z_2^2 \to \Z_2$, we cannot uniquely determine the $K$-theory group by the spectral sequence method.
However, we know from other methods that $KR^0(S^2) = \Z$ so that this differential must be surjective.
If in future research an explicit expression for the second differential is found, it would be interesting to compute it in this example.

\subsection{Time-reversal and a twofold rotation symmetry}\label{sec:rotation}

For a more exciting example, we now also include a rotation $R$ by $\pi$.
So we take the symmetry group $G = \Z_2 \times \Z_2 = \{1,R\} \times \{1,T\}$.
We twist the group so that the twisted group algebra satisfies the desirable physical situation on the quantum level, namely $R^2 = T^2 = -1$ and $TR = RT$. This thus represents spinful fermions on a two-dimensional square lattice with twofold rotation symmetry and hence the wallpaper group is $p2$. On the Brillouin torus $\mathbf{T}^2 = [-\pi,\pi]^2/\sim$ these symmetries act as $Tk = - k$ and $Rk = -k$.

Before the topological computations, we first have to compute the twisted Bredon coefficients, i.e. the representation rings and the relevant maps between them.
Note that the only stabilizers that occur are $G$ and $H := \{1,TR\}$, so we only have to compute twisted representations for these groups.
Because of this exceptional role played by $TR$ it is useful to set $S:= TR$ and forget about $T$ for the moment.
Note that in the twisted group algebra, $Si = - iS$, $S^2 = 1$ and $SR = RS$.
The twisted group algebras are abstractly isomorphic to matrix algebras:
\begin{align} \label{eq:algebra iso}
\leftidx{^\phi}{\C}{^{\tau}} H &= \frac{\R[i,S]}{(i^2 = -S^2 = 1,iS = -Si)} \cong |Cl_{1,1}| \cong M_2(\R)
\\ \label{eq:algebra iso2}
    \leftidx{^\phi}{\C}{^{\tau}} G &= \frac{\R[i,S,R]}{(i^2 = R^2 = -S^2 = 1,iS = -Si,RS = SR,iR = Ri)} \cong M_2(\C),
\end{align}
where the last isomorphism follows because the twisted group algebra is $\leftidx{^\phi}{\C}{^{\tau}} H \otimes_\R \C$.
Therefore the twisted group algebra of $H$ is Morita equivalent to the algebra $\mathbf{R}$, while the twisted group algebra of $G$ is Morita equivalent to the algebra $\mathbf{C}$.
The representation rings are therefore
\begin{align}
    ^\phi R^{\t-q}(H) & \cong KR^{-q}(\pt)
    \\
    ^\phi R^{\t-q}(G) &\cong K^{-q}(\pt),
\end{align}
see appendix \ref{sec: twisted reprings} for details on higher degree representation rings.
The restriction map in degree zero
\be
\Z \cong {}^\phi R^{\t}( G) \to {}^\phi R^{\t} (H) \cong \Z
\ee
is just given by mapping a complex vector space to its underlying real space and hence it is given by multiplication by two.
For $q = 1,$ the restriction map can only be zero, since $K^{-1}(\pt) = 0$.
For $q = 2$ restriction is a map 
\be
\Z \cong {}^\phi R^{\tau-2} (G) \to {}^\phi R^{\t-2}( H) \cong \Z_2,
\ee
so it can either be zero or reduction mod $2$. It is possible to explicitly check which it is by using explicit Clifford modules, but it turns out that we do not need to know which one it is in order to compute the $K$-theory.
The only remaining map between representation rings is the action of $R$ on the representation ring.
This action is given by conjugating modules over $\leftidx{^\phi}{\C}{^{\tau}} H$ with $R$.
Since $R$ is in the center of $\leftidx{^\phi}{\C}{^{\tau}} G$, the automorphism on $^\phi R^\t (H)$ resulting from this is trivial.
On the two relevant higher degree representation rings
\be
^\phi R^{\t-1}( H) = {}^\phi R^{\t-2}( H) = \Z_2
\ee
the action of $R$ is trivial as well because $\Z_2$ has no nontrivial automorphisms.

In order to compute the full twisted equivariant $K$-theory of the Brillioun zone torus, we first use the equivariant splitting method, giving the isomorphism \ref{eq: equivariantsplitting}.
Secondly we apply the spectral sequence on the components.
Note that the circles occuring in the isomorphism, i.e. $k_x = 0$ and $k_y = 0$ in the Brillouin zone, have identical group actions.
Hence they give isomorphic $K$-theory groups and we only have to compute one.
Next we have to decide on $G$-CW decompositions of our new spaces $S^2$ and $S^1$.
Since the action of $R$ is the same as the action of $T$, we can reuse the $G$-CW structure of the last example for $S^2$ as given in Figure \ref{fig:CW structure}.
For the circle we use the one-dimensional sub-$G$-CW complex of the $G$-CW structure on $S^2$.

Let us start by computing the twisted equivariant K-theory of the circle. We compute the $K$-theory by using \ref{eq: one-dimensional K-theory} for one-dimensional spaces.
The zeroth-cohomology $H^0_G(S^1,\leftidx{^\phi}{\mathcal{R}}{_G^{\tau}})$ is analogous to the example in the previous subsection. We can define  $\Z$-bases of equivariant $0$-cochains $\pi_0,\pi_\infty$ and $1$-cochains $\lambda$.
In contrast with the last example, we now have complex vector spaces on the fixed points and real vector spaces on the $k$-cells for $k>0$. 
Recall that the restriction map sends a complex vector space to its underlying real space and therefore this map is given by multiplication by two.
The Bredon differential is thus given by 
    \be
 d\pi_\infty(\ell) = \pi_\infty(\partial \ell)|_H = 2 \implies d\pi_\infty = 2\lambda.
    \ee
    Similarly  $d\pi_0 =-2\lambda.$
    Hence $H^0_G(S^1,\leftidx{^\phi}{\mathcal{R}}{_G^{\tau}}) \cong \Z $. Notice that for the first cohomology group $H^1_G(S^1,\leftidx{^\phi}{\mathcal{R}}{_G^{\tau-1}})$ the twisted representation ring of $G$ vanishes in te corresponding degree, so that the differential equals zero. Hence this cohomology group is equal to the group of equivariant $1$-cochains $\langle \lambda \rangle_{\Z_2}$, which equals the twisted representation ring of $H$ in degree $-1$. We conclude that $H^1_G(S^1,\leftidx{^\phi}{\mathcal{R}}{_G^{\tau-1}}) \cong \Z_2$. Via equation \eqref{eq: one-dimensional K-theory}, we arrive at
    \be
    \leftidx{^\phi}{K}{^{\tau}_G}(S^1) = \Z \oplus \Z_2.
    \ee
Since $\leftidx{^\phi}{K}{^{\tau}_G}(\pt) = {}^\phi R^\t( G) = \Z$, we see that $\leftidx{^\phi}{\widetilde{K}}{^{\tau}_G}(S^1) = \Z_2$ for both circles in the splitting of the torus.
These are precisely the invariants proposed by Lau et al. in \cite{PhysRevB.94.165164} and when non-trivial represent a M\"{o}bius twist in the Hilbert space of states along the invariant circles at $k_x = 0$ and $k_y = 0$. Our K-theory computation thus provides a mathematical proof of the existence of this invariant. 

Now we turn to the computation of the twisted equivariant $K$-theory of the 2-sphere. We use the same bases of equivariant cochains $\{\pi_0,\pi_\infty\}, \{\lambda\}$ and $\{\alpha\}$ as in the last example. For the zeroth cohomology, $H^0_G(S^2,\leftidx{^\phi}{\mathcal{R}}{_G^{\tau}})$, the computation is equivalent to the one in the previous subsection, hence $H^0_G(S^2,\leftidx{^\phi}{\mathcal{R}}{_G^{\tau}}) = \Z$. Going to $q = 1$, we see that there are no $0$-cochains, since $^\phi R^{\t-1}( G) =0$. The differential on $1$-cochains gives
    \begin{align}
    d\lambda(A) &= \lambda(\partial A)|_H = 
    \lambda(\ell) - \lambda(R \ell)|_H
    \\
    &= \lambda(l)|_H - R \lambda(\ell)|_H
    \\
    &= 0,
    \end{align}
    since $R$ necessarily acts trivially on $^\phi R^{\t-1}( H) = \Z_2$.
    Hence the cohomology groups are equal to the cochain groups:
    \begin{align}
        H^0_G(S^2,\leftidx{^\phi}{\mathcal{R}}{_G^{\tau - 1}}) = 0, \quad H^1_G(S^2,\leftidx{^\phi}{\mathcal{R}}{_G^{\tau - 1}}) = \Z_2, \quad 
        H^2_G(S^2,\leftidx{^\phi}{\mathcal{R}}{_G^{\tau - 1}}) = \Z_2.
    \end{align}
Since for $q = 2$ the $1$-cochains and $2$-cochains are exactly the same, the above computation also applies to the computation of the cohomology in degree $2$.
Therefore it follows that $H^2_G(S^2,\leftidx{^\phi}{\mathcal{R}}{_G^{\tau - 2}})$ equals $\mathbf{Z}_2$ as well. 
The relevant part of the second page is thus conveniently summarized in the following table.

\vspace{3mm}
\begin{tabular}{c|c|c|c}
& $p = 0$ & $p = 1$ & $p = 2$ \\
\hline
$q = 0$&$H^0_G(S^2,\leftidx{^\phi}{\mathcal{R}}{_G^{\tau}}) = \Z$    & & \\
\hline
$q = 1$  &$H^0_G(S^2,\leftidx{^\phi}{\mathcal{R}}{_G^{\tau - 1}}) = 0$   &$H^1_G(S^2,\leftidx{^\phi}{\mathcal{R}}{_G^{\tau - 1}}) = \Z_2$ & 
  $H^2_G(S^2,\leftidx{^\phi}{\mathcal{R}}{_G^{\tau - 1}}) = \Z_2$ \\
  \hline
$q = 2$     & & &$H^2_G(S^2,\leftidx{^\phi}{\mathcal{R}}{_G^{\tau - 2}}) = \Z_2$
\end{tabular}
\vspace{3mm}

The second differential $d_2: H^0_G(S^2,\leftidx{^\phi}{\mathcal{R}}{_G^{\tau}}) \to H^2_G(S^2,\leftidx{^\phi}{\mathcal{R}}{_G^{\tau - 1}})$ is either zero or reduction modulo $2$.
Independent of this distinction, the kernel of $d_2$ is abstractly isomorphic to $\Z$.
Hence the relevant part of the final page of the spectral sequence agrees with the diagonal $p = q$ in the table above.
The exact sequences \eqref{eq: exact sequences1} and \eqref{eq: exact sequences2} that follow from the spectral sequence now reduce to
\begin{align}
0 \to \Z_2 \to F \to \Z_2 \to 0,
\\
0 \to F \to \leftidx{^\phi}{K}{^{\tau}_G}(S^2) \to \Z \to 0.
\end{align}
Note that the second sequence splits.
Unfortunately, the first exact sequence implies only that $F = \Z_2^2$ or $F = \Z_4$.
Hence the Atiyah-Hirzebruch spectral sequence gives that $\leftidx{^\phi}{K}{^{\tau}_G}(S^2)$ is either $\Z \oplus \Z_2^2$ or $\Z \oplus \Z_4$, depending on whether the first exact sequence splits or not.
We can conclude from equation \ref{eq: equivariantsplitting} that 
\be
\leftidx{^\phi}{K}{^{\tau}}_G(\mathbf{T}^2) \cong \Z \oplus \Z_2^4 \quad \text{ or } \quad \leftidx{^\phi}{K}{^{\tau}}_G(\mathbf{T}^2) \cong \Z \oplus \Z_2^2 \oplus \Z_4.
\ee

To determine which of these two is the correct one, we employ the equivariant Mayer-Vietoris exact sequence. We can focus on the sphere since two possibilities for the $K$-theory originated there. Take open $G$-neighbourhoods of the north and south pole as $U_1 = S^2 \setminus \{p_\infty\}$ and $U_2 = S^2 \setminus \{p_0\}$.
Now consider the following part of the equivariant Mayer-Vietoris exact sequence with respect to $U_1$ and $U_2$:
\be
\dots \to {}^\phi K^{\tau - 1}_G (U_1 \cap U_2) \to {}^\phi K^\tau_G(S^2) \to {}^\phi K^\tau_G(U_1) \oplus {}^\phi K^\tau_G(U_2)  \to \dots
\ee
Note that both $U_1$ and $U_2$ are $G$-contractible to a point. 
Hence 
\[
{}^\phi K^\tau_G(U_1) \oplus {}^\phi K^\tau_G(U_2) \cong {}^\phi K^\tau_G(\pt) \oplus {}^\phi K^\tau_G(\pt) \cong \mathbf{Z}^2.
\]
Moreover, $U_1 \cap U_2$ is $G$-homotopy equivalent to the equator $S^1$.
To compute the twisted equivariant $K$-theory of this $S^1$, note that the action of the subgroup $H'$ generated by $R$ is free.
This implies that we can quotient this subgroup without having to worry about orbifold-type singularities.
Since there is a homeomorphism $S^1/H' \cong S^1$, we arrive at a circle with a trivial $H$-action.
Because twisted K-theory of orbifolds is invariant under equivalence \cite{gomi}, we see that
\be 
{}^\phi K^{\tau - 1}_G (U_1 \cap U_2) \cong {}^\phi K^{\tau - 1}_G (S^1) \cong {}^\phi K^{\tau - 1}_H (S^1).
\ee
The new twist is simply the restriction of the old twist to $H$ and since $(TR)^2 = 1$, the twist results in nonequivariant $KO$-theory.
Using suspensions and reduced $KO$-theory, we then arrive at
\begin{align}
{}^\phi K^{\tau - 1}_G (U_1 \cap U_2) &\cong KO^{-1}(S^1) \cong KO^{-1}(\pt) \oplus \widetilde{KO}^{-1}(S^1) 
\\
&\cong KO^{-1}(\pt) \oplus KO^{-2}(\pt) \cong \mathbf{Z}_2^2.
\end{align}
Hence the equivariant Mayer-Vietoris sequence takes on the form
\be\label{MVpart}
\dots \longrightarrow \mathbf{Z}_2^2 \overset{f_1}{\longrightarrow} {}^\phi K^\tau_G(S^2) \overset{f_2}{\longrightarrow} \mathbf{Z}^2  \longrightarrow \dots
\ee
A simple diagram chasing argument now shows that ${}^\phi K^\tau_G(S^2)$ does not contain any $4$-torsion.
Indeed, suppose that $a \in {}^\phi K^\tau_G(S^2)$ satisfies $4a = 0$. 
We will show that this implies $2a = 0$.
First note that $4 f_2(a) = f_2(4a) = 0$.
However, the image of $f_2$ is torsion-free so we necessarily have that $f_2(a) = 0$.
We can conclude that $a$ is in the kernel of $f_2$.
By exactness, this implies that there is some $b \in \mathbf{Z}_2^2$ such that $f_1(b) = a$.
Now it follows by the group structure of $\mathbf{Z}_2^2$ that $2a = 2f_1(b) = f_1(2b) = f_1(0) = 0$ as desired. 

The twisted equivariant $K$-theory for $p2$ symmetry in class AII is thus,
\be
\leftidx{^\phi}{K}{^{\tau}}_G(\mathbf{T}^2) \cong \Z \oplus \Z_2^4,
\ee
which exactly agrees with the heuristic arguments proposed in \cite{Kruthoff2}. In particular, there it is argued that each $\Z_2$ invariant comes from a vortex anti-vortex pair in the Berry connection stuck at the high-symmetry points. In this case, there are four of such point and hence four $\Z_2$ invariants. The part of this $K$-theory coming from the sphere was also computed in \cite{PhysRevB.90.165114}. Our method now provides a rigorous mathematical proof of this computation. 

Notice that the spectral sequence also gives insight on the origin of the invariants.
First of all, the $\Z$ factor is simply the rank of the bundle. 
Next we have two $\mathbf{Z}_2$'s that are M\"obius-type line invariants along the two cycles of the torus and are simply the invariants already found in \cite{PhysRevB.94.165164}. 
On the Brillouin zone sphere remaining after the equivariant splitting, we have again one such $\mathbf{Z}_2$-line invariant.
Finally, we have the fourth $\mathbf{Z}_2$-invariant, which is a Fu-Kane-Mele-type surface invariant on the Brillouin sphere. We discuss a possible connection between these invariants and the vortex picture in the Section \ref{discussion}.
Also note that from the equivariant Mayer-Vietoris exact sequence we could only determine the type of torsion and not the full $K$-theory group. It is the combination with the Atiyah-Hirzebruch spectral sequence that resulted in the final answer. 

\section{Generalizations}\label{sec:generalisations}

We will now discuss various generalizations of the simple examples we studied in the previous section. Furthermore, we will give an algorithmic method to compute the twisted representations rings.

\subsection{Other crystal symmetries}\label{Othersymm}
In the above we performed computations involving either no spatial symmetries or a twofold rotation. The computation of the last section can be generalized straightforwardly to a fourfold rotation.
The only thing that requires extra attention is that even though the splitting \eqref{eq: equivariantsplittingb} holds true, the $K$-theory does not split further according to equation \eqref{eq: equivariantsplitting}.
This is because the fourfold rotation interchanges the two circles of the figure eight.
So one really has to compute the K-theory of the figure eight.
The final result for a fourfold rotation in class AII is 
\be\label{p4resultAII}
{}^{\phi}K^{\t}_{\mathbf{Z}_4 \times \mathbf{Z}_2^T}(\mathbf{T}^2) = \mathbf{Z}_2^3 \oplus \mathbf{Z}^3.
\ee
Just as for a twofold rotation, there are two torsion invariants coming from the spherical Brillouin zone: one is a line invariant and one a surface invariant.
However, there is only one $\mathbf{Z}_2$-invariant corresponding to the boundary figure eight, since the two line invariants of the last section are identified by the fourfold rotation.
It is amusing to see that this computation exactly agrees with the more heuristic arguments presented in \cite{Kruthoff2}. There are three high-symmetry points (not related by any symmetry operation) at which vortex anti-vortex pairs can be stuck, giving rise to three $\Z_2$ invariants. The free part of the $K$-theory is simply the consistent assignment of representations as discussed in previous subsections. 

It is interesting to do the same computation for wallpaper groups with reflections. For instance, let us consider a two-dimensional crystal with a single reflection symmetry. On the Brillouin torus this symmetry acts as $t \cdot k = (k_x,-k_y)$ for $k \in \mathbf{T}^2$. Just as for the twofold rotation symmetry, we have that on the fibers $t^2 = -1$. Going through the spectral sequence analysis, we find that the spectral sequence method gives a unique answer by itself.
The $K$-theory is given by 
\be\label{reflectionresultAII}
{}^{\phi}K^{\t}_{\mathbf{Z}_2 \times \mathbf{Z}_2^T}(\mathbf{T}^2) = \mathbf{Z}_2^2 \oplus \mathbf{Z},
\ee
in class AII. This result also agrees with \cite{Kruthoff2}.  The $\mathbf{Z}_2$ invariants come from the circles fixed under the reflection symmetry.
Note that one of these circles is part of the figure eight boundary, whereas the other comes from the circle of reflection on the sphere after the equivariant splitting \eqref{eq: equivariantsplittingb}. The part of the $K$-theory coming from the sphere (in particular exactly one of the two $\Z_2$ invariants) was also obtained in \cite{PhysRevB.88.075142, PhysRevB.90.165114, PhysRevB.88.125129}. 

For more complicated symmetries, the method becomes rather involved. 
For example, symmetries with elements of odd order can in two spatial dimensions only occur on a hexagonal lattice, so one has to take the non-trivial identifications of the Brillouin zone into account. The Brillouin zone will still be a torus, but the equivariant splitting we used in the previous section becomes more difficult. 
Also we cannot make use of the real structure $TR$ in the same way we did above; in case of an odd order rotation the stabilizer of a generic point will be trivial instead.
Moreover, $TR$ will anticommute with reflections in class AII, see the example at the end of this section.
The basic difficulty as the symmetry group gets larger is computing the twisted group algebras and their representation rings; determining them using abstract algebra as we did in equations \eqref{eq:algebra iso} and \eqref{eq:algebra iso2} quickly becomes tedious. One way forward is to give an alternative way of describing and constructing the twisted representation rings. Essentially, the only requirement for the construction of these twisted representation rings is knowing how many representations exist and which are real, complex and quaternionic. More precisely, if the number of real, complex and quaternionic irreducible $(\phi,\tau)$-twisted representations of $G$ is denoted by $n_k$ with $k = \mathbf{R},\mathbf{C}, \mathbf{H}$, we have  
\be
^\phi R^\t (G) = K^0(\pt)^{n_{\mathbf{C}}} \oplus KR^0(\pt)^{n_{\mathbf{R}}} \oplus KR^{-4}(\pt)^{n_{\mathbf{H}}}.
\ee
Once this data has been computed, the higher order rings follow from the Bott clock, see appendix \ref{sec: twisted reprings}. Remember that although we refer to $^\phi R^\t (G)$ as rings, they are actually not rings. The task we are thus left with is to determine the integers $n_k$. In other words, we need an adequate representation theory for twisted groups. From a physics point of view, such a theory was outlined in \cite{bradleycracknell} and we will now showcase this to make contact with the approaches to such problems by the crystallography community. It would also be interesting to see this point of view compared with the Wigner test, which is the appropriate generalization of the Frobenius-Schur indicator as given in \cite{bradleycracknell}. Although we will not give a rigorous proof of the connection between the representation theory of space groups and twisted representation rings, we will give evidence for such a connection below and in appendix \ref{sec: twisted reprings}.

In the rest of this section, we will mention the procedure of \cite{bradleycracknell} and illustrate it using a simple example. For this we will need to briefly change gears. We formulate twisted groups in terms of double covers and twisted representations as double-valued representations. To see how this formulation is related to the twists $\t$ in the main text, the reader may wish to consult the final paragraph of appendix \ref{sec:Freed Moore twists}. Although this procedure works for both class AI and AII, let us focus on the latter. For simplicity we assume that $G = G_0 \times \mathbf{Z}_2^T$, where $\mathbf{Z}_2^T$ represents the time-reversal action on the Brillouin zone and $G_0$ consists of the other symmetries. We have written $G$ in this way, because $G_0$ will be lifted to a linear action on the Hilbert space, whereas time-reversal lifts to an anti-linear action. Let us focus first on $G_0$. In class AII, we are dealing with fermions (class AI assumes the system is bosonic) and we have to consider a certain double cover of $G_0$, which we denote by $\widehat{G}_0$. The representations of the double cover take the usual signs into account that come for example from rotations over $2\pi$ acting with a minus sign on the Hilbert space.
In this sense they form the structure analogous to the twist $\tau$ in the rest of this paper.
It is usually intuitively clear which double cover is desirable, but for a general point group $G_0$ of a $d$-dimensional lattice it can be described by the following abstract mathematical construction.
The double cover should be the pullback of the negative Pin group $\operatorname{Pin}_-(d)$ covering the orthogonal group:
\begin{center}
\begin{tikzcd}
    1  \arrow[r]
    &\Z_2 \arrow[d,equal] \arrow{r}
    & \widehat{G}_0 \arrow[d,hook]\arrow[r]
    & G_0 \arrow[d,hook]\arrow[r]
    & 1
    \\
    1 \arrow[r]
    & \Z_2\arrow[r]
    &  \operatorname{Pin}_-(d) \arrow[r]
    & O(d)  \arrow[r]
    &  1.
    \end{tikzcd}
\end{center}
The reason that it is not the spin group is that this group would only account for rotations, not for reflections.
In other words, $\operatorname{Spin}(d)$ is only a double cover of $SO(d)$. The reason that we pick the negative Pin group $\operatorname{Pin}_-(d)$ instead of the other central extension $\operatorname{Pin}_+(d)$ of $O(d)$, is that in $\operatorname{Pin}_+(d)$ reflections square to the identity instead of to the desired $-1$\footnote{When other symmetries such as gauge or flavour symmetries are present or when there are interactions, the double cover could also be $\operatorname{Pin}_-(d)$. Although it would be interesting to map out all possible choices, it is beyond to scope of the present work.}.
It is useful to note that if $G_0 = D_n$ is the symmetry group of an $n$-gon, then $\widehat{G}_0$ is known as the dicyclic group of order $4n$.
Twisted representations of the group $G_0$ can now be described as certain ordinary representations of the double cover group $\widehat{G}_0$.
Let us call representations fermionic when the newly introduced subgroup $\Z_2 \subseteq \widehat{G}_0$ acts nontrivially.
Representations where the $\Z_2$ acts trivially are called bosonic.
This exactly means that $2\pi$ rotations (or double reflections) act by $-1$ on fermionic representations and trivially on bosonic representations. In general, the double cover will admit both fermionic and bosonic representations, but for systems in class AII the Hilbert space is organized in terms of fermionic representations only. Given these fermionic representations, we are now ready to add in time-reversal symmetry. This enlarges the group with another generator $T$ that squares to minus one. Set-theoretically, we can write the full group acting on the Hilbert space as \footnote{Note that in the previous section, we used $G$ to denote the group acting on the Brillouin zone. The group $\widehat{G}$ is the lift of that group to a group acting on the Hilbert space in which time-reversal, $2\pi$ rotation and double reflections square to minus one. Hence fermionic representations of $\widehat{G}$ are equivalent to twisted representations of $G$.}
\be\label{fullgroup}
\widehat{G} = \widehat{G}_0 \sqcup A \widehat{G}_0,
\ee
where $A$ is an antiunitary symmetry operator, i.e. $\phi(Ag) = -1$ for any $g \in \widehat{G}_0$. Notice that starting with any group $\widehat{G}$ and any nontrivial homomorphism $\phi: \widehat{G} \to \Z_2$, we could have created such a decomposition by taking $\widehat{G}_0 = \ker \phi$ and picking some $A \notin \ker \phi$. Usually one takes $A = T$, but other forms of $A$ are also possible. For example, in the example in the previous subsection, for $H :=\{1,TR\}$ we would have $A = TR$ and $\widehat{G}_0 = \{1,-1\}$, with $-1$ the non-trivial element in the double cover of the trivial group. 
Notice also that the choice of $A$ in \eqref{fullgroup} is a bit arbitrairy as $A' = Ag$ for $g\in \widehat{G}_0$ instead of $A$ will give rise to unitarily equivalent representations. 

Let us denote a fermionic representation of $\widehat{G}_0$ by $\rho$, which we can assume to be unitary. As shown in \cite{bradleycracknell}, the fermionic (matrix) representations $D$ of $\widehat{G}$ are then given by 
\be\label{Dg}
D(g) = \begin{pmatrix}
\rho(g) & 0 \\
0 & \bar{\rho}(A^{-1}gA)
\end{pmatrix}
\ee
and 
\be\label{DAg}
D(Ag) = \begin{pmatrix}
0 & \rho(AgA) \\
\bar{\rho}(g) & 0
\end{pmatrix}
\ee
for $g \in \widehat{G}_0$. Here $\bar{\rho}$ denotes the complex conjugate of the representation $\rho$. When $A$ is just $T$, we can commute it with $g \in \widehat{G}_0$ and the expressions become much simpler. In general we can always write $A$ as $T$ times something in $G_0$ and then it will appear quadratically or not at all. Hence $T$ can be factored out by using the fact that $\rho$ is a homomorphism. Thus, the representation $D$ will only depend on the sign of $T^2$, which for our case is minus one. For these representations the notion of reducibility is similar to ordinary representations, see \cite{bradleycracknell}. 

Now there are three different cases to distinguish: $a)$ Either time-reversal symmetry does nothing to the representation, $b)$ two unitarily equivalent representations of dimension $k$ form a new (irreducible) representation of dimension $2k$, or $c)$ complex conjugate irreducible representations of dimension $l$ form a representation of dimension $2l$. These three cases can be described as follows: 
\begin{enumerate}
\item[a)] In this case $\rho(g)$ is unitarily equivalent to $\bar{\rho}(A^{-1}gA)$, i.e. $\rho(g) = N\bar{\rho}(A^{-1}gA)N^{-1}$ for some fixed unitary matrix $N$ and $g \in \widehat{G}_0$. Moreover, $N$ satisfies $N\bar{N} = +\rho(A^2)$, and then $D(g) = \rho(g)$ and $D(Ag) = \pm  \rho(AgA^{-1})N$ \footnote{In this case, the matrix representations of $g$ and $Ag$ for $g \in \widehat{G}_0$ given in \eqref{Dg} and \eqref{DAg} can both be made block diagonal and in fact the representation $D$ is reducible. Consequently, $\rho$ and $D$ have the same dimensionality. The $\pm$ appearing for $D(Ag)$ represents to unitary equivalent representations. See \cite{bradleycracknell} for more details.}.
\item[b)] In this case $\rho(g)$ is unitarily equivalent to $\bar{\rho}(A^{-1}gA)$, i.e. $\rho(g) = N\bar{\rho}(A^{-1}gA)N^{-1}$ for some fixed unitary matrix $N$ and $g \in \widehat{G}_0$. However, $N$ satisfies $N\bar{N} = -\rho(A^2)$, and then 
\be
D(g) = \begin{pmatrix}
\rho(g)  & 0 \\
0 & \rho(g)
\end{pmatrix},
\quad  D(Ag) = \begin{pmatrix}
0 & -\rho(AgA^{-1})N  \\
\rho(AgA^{-1})N & 0 
\end{pmatrix}.
\ee
\item[c)] In this case $\rho(g)$ is not unitarily equivalent to $\bar{\rho}(A^{-1}gA)$. The representations are then given by
\be
D(g) = \begin{pmatrix}
\rho(g)  & 0 \\
0 & \bar{\rho}(A^{-1}gA)
\end{pmatrix},
\quad  D(Ag) = \begin{pmatrix}
0 & \rho(AgA) \\
\bar{\rho}(g) & 0 
\end{pmatrix}.
\ee
\end{enumerate}
It is clear that representations of type $c)$ always correspond to complex representations. Type $a)$ are the real representations and type $b)$ are quaternionic representations. An important subtlety is when the unbroken symmetry group of the fixed point does not contain any antiunitary symmetries. In that case the representations remain complex, just as we saw in Table \ref{TwistedRing} in the case of a trivial stabilizer group. We will now study some simple examples to see how this works in practice. 

For a rotation symmetry $\Z_n$, the double cover is $\Z_{2n}$ which has $n$ complex fermionic representations for $n$ even. For $n$ odd, there are $n-1$ complex fermionic representations and one quaternionic representation. Due to time-reversal symmetry the complex fermionic representations will pair up and hence the twisted representation ring of degree $-q$ of $G = \Z_n \times \Z_2^T$ is
\be
{}^{\phi}R^{\t - q}(G) = \left\{ \begin{array}{cc}
K^{-q}(\pt)^{n/2} & \text{ if } n = \text{even }\\
K^{-q}(\pt)^{(n-1)/2} \oplus KR^{-q-4}(\pt) & \text{ if } n = \text{odd }
\end{array}\right. .
\ee 
This is exactly the same result as one would get by constructing the representation ring of the twisted group algebra, which is a more formal way of computing the twisted equivariant $K$-theory of a point. More details on that construction can be found in the appendix. 
A more non-trivial example is $G_0 = \mathbf{Z}_2 \times \mathbf{Z}_2$. This group is generated by $t_1$ and $t_2$, which represent reflections in the $k_x$ and $k_y$ axis respectively. The double cover of this group is $Q_8$, the quaternion group. The action of $G_0$ on the Brillouin zone torus $\mathbf{T}^2$ has four fixed points $(0,0)$, $(0,\pi)$, $(\pi,0)$ and $(\pi,\pi)$ and four fixed circles $(0,k_y)$, $(\pi,k_y)$, $(k_x,0)$ and $(k_x,\pi)$. The fixed points have stabilizer group $G_0$. We will analyze the representations of the cover of this group first. The group $Q_8$ has five representations of which only one is fermionic, since for all other representations $-1 \in Q_8$ acts trivially. This fermionic representation is just its regular action on the quaternions $\mathbf{H}$, which is a two-dimensional representation over the complex numbers. We denote the generators of $Q_8$ by $\hat{t}_1$ and $\hat{t}_2$. The representation is concretely given by 
\be
\rho(\hat{t}_1) = i\s_1, \quad \rho(\hat{t}_2) = i\s_2
\ee
where $\s_i$ are the Pauli matrices. Note that since we are at fixed points whose stabilizer group is the full point group, we have $A = T$. To determine what time-reversal does with these representations, we have to find out whether $\rho$ and $\bar{\rho}$ are unitarily equivalent. Clearly this is the case, since $N = i\s_2$ is an explicit unitary matrix that intertwines $\rho$ with $\bar{\rho}$. To see this, note that it anticommutes with all purely imaginary matrices in this representation. Given this $N$, we have
\be
N\bar{N} = -\s_2^2 = -1 = +T^2.
\ee
Thus we are in case $a)$ and we have $^\phi R^{\tau - q}(G) = KR^{-q}(\pt)$. 

For the fixed circles the twisted representation theory of the stabilizer is unexpectedly interesting. Consider for example the circle $k_y = 0$. This circle is fixed by $H = \{1,t_2\}$, which lifts to the double cover $\widehat{H} = \{1,\hat{t}_2, \hat{t}_2^2, \hat{t}_2^3\}$. The full set of elements in the double cover (including time-reversal) that leave $k_y = 0$ fixed is $\{1,\hat{t}_2, \hat{t}_2^2, \hat{t}_2^3\} \sqcup TR\{1, \hat{t}_2, \hat{t}_2^2, \hat{t}_2^3\}$, hence we pick $A = TR$, where $R = \hat{t}_1 \hat{t}_2$ is the lift of the twofold rotation in the double group. Note that $A^2 = 1$, but since $A$ contains the rotation $R$ it anticommutes with the reflections. Even though the fermionic representations $\rho_{\pm}$ of $\widehat{H}$ have complex eigenvalues $\pm i$ for the reflection, these two irreps nevertheless belong to case $a)$. To see this, first note that since these representations are one-dimensional over the complex numbers, $N$ drops out. Hence two such representations are unitarily equivalent if and only if they are equal. Now note that there is an extra minus sign that cancels the minus sign coming from complex conjugation. Indeed we have $\bar{\rho}_\pm = \rho_{\mp}$ and hence
\be 
\bar{\rho}_{\pm} (A^{-1}\hat{t}_2 A) = \rho_{\mp}(-\hat{t}_2) = \rho_{\pm}(\hat{t}_2).
\ee
For the other fixed circles the exact same argument holds. At the fixed circles we therefore have two real representations and hence the representation ring is
\be
^\phi R^{\t - q}(H \times \mathbf{Z}_2^T) = KR^{-q}(\pt) \oplus KR^{-q}(\pt).
\ee
From this computation, one immediately sees that a lot of torsion will appear in the spectral sequence. We will not compute the full K-theory for this crystal group here, but we expect the exact sequences that result from the spectral sequence to split. 

The approach given in this section has a natural extension to non-symmorphic symmetries, but the computations become more tedious. We will discuss these symmetries further in the discussion section.

\subsection{Class AI}

Topological insulators in class AI satisfy $T^2 = 1$. The bundle therefore has a real structure given by $T$. In the absence of a point-group symmetry, the $K$-theory classifying topological insulators is Real $K$-theory $KR$, which we computed in \eqref{K-theory of 2-torus}. When the topological insulator has a non-trivial crystal symmetry, the twist $\t$ is trivial. Therefore we have to compute equivariant Real $K$-theory, which for a group $G$ is denoted by $KR_G$. We can again use the spectral sequence method to compute the relevant $K$-theory groups. For a single reflection symmetry in class AI we find that
\be
KR_{\Z_2}(\mathbf{T}^2) = \Z^3,
\ee
which is again in agreement with \cite{PhysRevB.88.075142, PhysRevB.90.165114, PhysRevB.88.125129}. The invariants are purely coming from the representations at the fixed points together with a non-trivial glueing condition. 

It turns out that for other symmetries, the $K$-theory for class AI is much harder to compute. In the examples we considered, the computations are plagued by the higher differentials of the spectral sequence. This problem appears especially in class AI because in two dimensions, the second differential can only make a nontrivial contribution in case the (twisted) stabilizer $H_k$ of some zero-cell $k$ admits a real representation. In contrast to class A (where all representations are complex) and in class AII (where real representations do appear sometimes), real representations are the norm in class AI.
This is because time-reversal is a real structure.
In particular whenever $H_k$ contains $T$, the trivial representation of $H_k$ is always real.
Therefore the second differential can often make contributions to the $K$-theory, but we have not been able to show what these contributions are for a general crystal symmetry. What we do know however is that for the analogous computation of the one done in \ref{sec:rotation} (i.e. time-reversal and a twofold rotation symmetry) the $K$-theory is one of the two groups
\be
KR_{\Z_2}(\mathbf{T}^2) =  \mathbf{Z}^5\text{ or } KR_{\Z_2}(\mathbf{T}^2) = \mathbf{Z}^5 \oplus \mathbf{Z}_2 .
\ee
We were not able to show which of these two is the actual answer, but the analysis in \cite{PhysRevLett.106.106802} suggests that there cannot be any torsion in this case. In this work a fourfold rotation symmetry is considered and it is shown that a $\Z_2$ invariant appears because the two complex representations form a two dimensional representation at the fixed points $(0,0)$ and $(\pi,\pi)$ once time-reversal symmetry is taken into account. An effective time-reversal operator $TR$ can be defined that squares to minus one and hence at these points the vector bundle has the same quaternionic structure that is found in class AII. This observation was crucial to show that there is a single $\Z_2$ for $p4$ in class AI. For a twofold rotation symmetry a similar construction does not work, which is why we believe $KR_{\Z_2}(\mathbf{T}^2) =  \mathbf{Z}^5$. From the point of view of the spectral sequence, this would be the case if the second differential $d_2$ (of which no convenient explicit expression is known) is surjective. In \cite{Shiozaki:2018srz}, it is indeed argued that in this case $d_2$ is surjective, just like we found in case when no point symmetries were present. 

We have also computed the KR-theory associated to $p4$ rotation symmetry in class AI, but we ran into the same problems as for $p2$. However, from the above discussion, we know that there should be a single $\Z_2$ invariant. If a closed expression for the second differential is derived, it would therefore be interesting to show rigorously that $d_2$ is surjective for $p2$, but not surjective for $p4$.

In case of more non-trivial crystal symmetries, the representation theory and hence its twisted rings can be computed using the technique outlined above. The difference is that now we are interested in the bosonic representations of $\hat{G}$. In other words, we do not have to consider the double cover group, but instead we can work with the point group itself, thus \eqref{fullgroup} changes to $G = G_0 \sqcup AG_0$.
If $A$ happens to commute with all $g \in G_0$, this means we just have to determine the real representation theory of $G_0$. So, for example let us consider a point group $D_4$. This group has five representations and all are realizable over the real numbers. These representations are thus in case $a)$. The twisted representation ring is then 
\be
^\phi R^{\t - q}(D_4 \times \Z_2^T) = KR^{-q}(\pt)^5
\ee
Again for a generic point on the torus, there is an unbroken symmetry group $H = \{1,TR^2\}$, hence a smart choice would be to pick $A = TR^2$. Notice that again $(TR^2)^2 = 1$ and $TR^2$ commutes with all elements of $G$. For the subgroup $H$ we have $H_0 = \{1\}$, whose representation obviously belongs to case $a)$, hence ${}^\phi R^{\tau - q} (H) = KR^{-q}(\pt)$. The geometric action of $D_4$ on the torus has three other non-trivial stabilizer groups, two isomorphic to $\Z_2$ and one isomorphic to $\Z_2 \times \Z_2$. The representation rings for these two groups also only consist of copies of $KR^{-q}(\pt)$, since all representations are real. In fact, we have 
\begin{align}
^\phi R^{\t - q}(\Z_2 \times \Z_2^T) = KR^{-q}(\pt),\\
^\phi R^{\tau - q}(\Z_2 \times \Z_2 \times \Z_2^T) = KR^{-q}(\pt).
\end{align}

\subsection{Class A}
In class A the spectral sequence is well-known \cite{dwyertwisted,barcenasspectral} and the computations are a lot more tractable. There are no antiunitary operators and the $K$-theory is just Atiyah \& Segal's complex equivariant $K$-theory \cite{segalequivariant}, but possibly twisted. The twist now comes purely from non-symmorphic symmetries. Let us focus on symmorphic symmetries so that there is no twist. In this case, the relevant exact sequences in the spectral sequence constructed here will always split and so unlike in class AI and AII, we get a unique answer. One might wonder whether the higher differentials could give non-trivial contributions. Firstly, since $\mathcal{R}^{-q}_{G} = 0$ for odd $q$, every other row on the second page of the spectral sequence is trivial so that the second differential always vanishes. 
Moreover, in two dimensions the third and higher differentials always vanish. Thus we can easily determine the equivariant $K$-theory exactly.

Indeed, since in the complex case $\mathcal{R}^{-2}_G = \mathcal{R}_G$, one quickly sees that the exact sequences \eqref{eq: exact sequences1} and \eqref{eq: exact sequences2} imply that
\be \label{KtheoryA}
K_G(X) \cong H^0_G(X,\mathcal{R}_G) \oplus H^2_G(X,\mathcal{R}_G).
\ee 
This isomorphism also holds in three dimensions, since in that case the third differential gives no additional contribution.
We will illustrate this fact by a short argument.
If $X$ is three-dimensional, the arguments above give us the isomorphism
\be
K_G(X) \cong \ker \left( d_3:  H^0_G(X,\mathcal{R}_G) \to H^3_G(X,\mathcal{R}_G)\right) \oplus H^2_G(X,\mathcal{R}_G).
\ee 
Now note that in class A, the Bredon cochains map into ordinary representation rings of subgroups $H \subseteq G$.
Since these representation rings are always torsion-free, so are all groups of Bredon cochains. Since $H^0_G(X,\mathcal{R}_G)$ is a kernel of a $\Z$-linear map between Bredon cochains, it must therefore also be torsion-free and hence so is $\ker d_3$.
But by the equivariant Chern character isomorphism \cite{cherncharacter}
\be
K_G(X) \otimes \mathbf{C} \cong H^0_G(X,\mathcal{R}_G \otimes \C) \oplus H^2_G(X,\mathcal{R}_G \otimes \C),
\ee
we already know that the formula \eqref{KtheoryA} holds modulo torsion. Therefore, there must be an isomorphism $\ker d_3 \cong H^0_G(X,\mathcal{R}_G)$. Note that even though this argument does not imply that $d_3$ vanishes, it still implies that it can be ignored in abstract computation.

Let us reflect on the results we have just established in class A. First of all, in light of \cite{Kruthoff1}, we see that there is indeed a clear distinction between the representations at fixed points and how they are glued to the representations at lines on the one hand, which are captured by $H^0_G(X,\mathcal{R}_G)$, and higher-dimensional invariants, such as Chern numbers on the other hand, which are in $H^2_G(X,\mathcal{R}_G)$. In two dimensions, one can check by explicit computation that $H^2_G$ is torsion-free, but in three dimensions it is known that it contains torsion in certain examples \cite{Shiozaki:2018srz}. Nevertheless, the torsion-free part is still captured by the proposed algorithm in \cite{Kruthoff1}. The torsion is hard to understand systematically, but intuitively one expects it to arise from either non-symmorphic space groups or non-trivial identifications due to the crystal structure. Since the Bredon cohomology of a complex is something purely combinatorial, equation \ref{KtheoryA} provides an algorithmic approach to computing the $K$-theory for class A in full generality. It would be interesting to develop such an algorithm and compare it to the results of \cite{Shiozaki:2018srz}. In fact, comparing with existing literature on the equivariant K-theory associated with the space group $F222$, we see that \cite{Shiozaki:2018srz} obtains an $\Z_2$ invariant in class $A$, whereas in \cite{mcalister2005noncommutative} this K-theory is found to be torsion-less. We leave a detailed analysis of this discrepancy to future work.

We also briefly mention a useful alternative method to determine the $K$-theory groups in class A in some cases. 
Namely, in the cases in which the equivariant splitting method applies, the classification is determined by the equivariant $K$-theory of representation spheres.
The nontwisted complex $K$-theory of representation spheres is easily determined in terms of purely representation-theoric data as was described by Karoubi, see the survey paper \cite{representationspheres}.
It would be interesting to research whether pure representation-theoretic data could describe the K-theory of representation spheres in case time-reversal is included.

\section{Discussion}\label{discussion}

We have outlined a way to compute the $K$-theory that classifies topological insulators with or without time-reversal symmetry and with non-trivial crystal symmetry. Using an Atiyah-Hirzebruch spectral sequence, we computed these groups in a couple of examples and saw that often more work is needed to compute the exact answer. Nevertheless, it is noteworthy that the classification with $K$-theory matches with the rather heuristic arguments presented in \cite{Kruthoff1, Kruthoff2}, at least for the examples that we computed. Moreover, with the techniques of crystallography, we were able to give an algorithmic way of computing the twisted representation rings in any degree. 

In our $K$-theory computations, we have also stumbled upon some difficulties that in general seem hard to overcome.
Firstly, there is the fact that as of yet no explicit expression for the higher differentials of the spectral sequence is known, even in the simplest cases.
For nonequivariant complex $K$-theory, it is known that the second differential vanishes and the third differential is the extended third Steenrod square $Sq^3$, which is the composition
\be
H^p(X,\mathbf{Z}) \to H^p(X,\mathbf{Z}_2) \overset{Sq^2}{\to} H^{p+2}(X,\mathbf{Z}_2) \overset{\beta}{\to} H^{p+3}(X,\mathbf{Z}),
\ee
where $Sq^2$ is the second Steenrod square and $\beta$ is the Bockstein homomorphism associated to the exact sequence
\be
0 \to \Z \overset{\times 2}{ \to} \Z \to \Z_2 \to 0.
\ee
This result has been generalized to twisted complex $K$-theory in \cite{twistedktheory}, but even in nontwisted equivariant $K$-theory the situation is much more involved as is illustrated in \cite{barcenasspectral}.
Real $K$-theory on the other hand even introduces a second differential and is less studied in the literature.
For $KO$-theory (the $K$-theory that classifies real vector bundles instead of complex ones, i.e. $KR$-theory with trivial involution) it has long been known that the second differential is the (appropriately extended) second Steenrod square \cite{seconddifferentialKO}.
Keeping the applications in mind, it would be interesting to work out explicit expressions of the higher differentials for small groups and CW-complexes from their abstract definition.
Intuitively, a non-trivial $r$th differential represents obstructions of extending the vector bundle on a $d$ dimensional subspace to an $d + r$ dimensional subspace. This intuitive understanding was used in \cite{Shiozaki:2018srz} to argue that a non-trivial $r$th differentials is an obstruction of smoothly extending (i.e. without gap closing) a topological insulator on a $d$-cell to an $(d+r)$-cell. Surprisingly, this allowed the authors to construct explicit expressions for the higher differentials in specific examples. It would be would be interesting to rigorously show that this construction works in general.

The second and more fundamental difficulty in using the spectral sequence method is the problem of non-unique extensions.
Indeed, since the exact sequences \eqref{eq: exact sequences1} and \eqref{eq: exact sequences2} are not always split, we cannot determine the $K$-theory uniquely unless we explicitly know the maps involved.
In Section \ref{sec:rotation} we had to face this problem, since torsion groups appeared both in degree 1 and degree 2 simply because $KR^{-1}(\pt) = KR^{-2}(\pt) = \Z_2$.
With just the Atiyah-Hirzebruch spectral sequence in our toolbox, this problem could only be solved by explicitly determining all maps involved in our exact sequences, which is tedious even in simple examples.
In order to fully determine the (especially 2-)torsion invariants for general point groups, we will need a supplement to the spectral sequence.
The supplement we used in Section \ref{sec:rotation} was the equivariant Mayer-Vietoris exact sequence.
There are several other possibilities for such a supplement. One would be an Adams-type spectral sequence. Such spectral sequences are made precisely to measure the torsion part of groups of stable homotopy classes of maps between spaces. Another supplement, which was recently discussed in \cite{Gomi:2018uhs}, relates a K-theory with a non-unique extension problem to one which does have a unique extension through a notion of T-duality. 


Setting aside these difficulties, it would also be interesting to apply our method to topological superconductors and insulators with a chiral symmetry. These cases cover the remaining $7$ Altland-Zirnbauer classes \cite{altlandzirnbauer} and might also give many new invariants that can be studied experimentally. A simple example would be to study a topological superconductor with only particle hole symmetry $C$ which can square to $+1$ or $-1$. Such systems are in class $D$ and $C$, respectively. In two dimension the classification without any symmetry is just $\mathbf{Z}$ and it would be interesting to see how crystal symmetry changes this.
However, even though chiral symmetries are incorporated in the framework of Freed \& Moore's $K$-theory, it does not seem to be well-suited for this purpose.
For example, for class AIII a short argument shows that the Freed-Moore $K$-theory group of $S^1$ vanishes in case no other symmetries are present.
This contradicts the ten-fold way, which says that in one dimension class AII topological insulators on a spherical Brillouin zone are classified by $\Z$.
As argued in \cite[\S 3.5]{gomi}, this discrepancy results from the fact that the types of $K$-theory that include chiral symmetries are no longer realizable by finite-dimensional bundles as Freed and Moore assume.
However, this seems to contradict the physical principle that our topological insulators only admit a finite number of bands.
Assuming that the $K$-theory defined in \cite{gomi} is the physically relevant type of $K$-theory, we know from this work that it satisfies the desired axioms for cohomology. 
In that case, the higher representation rings admit an obvious generalization to particle-hole reversing symmetries and a version of the spectral sequence similar to the one developed here therefore probably holds.

Another very interesting class of symmetries, which we have not touched upon yet, are non-symmorphic symmetries. In two dimensions most symmetries are symmorphic, but in three dimensions there is a large class of crystals that exhibit some form of non-symmorphicity. The implementation of such symmetries in our recipe is mathematically challenging, because non-symmorphic crystals give twists that can vary throughout the Brillouin zone. The known representation theory of non-symmorphic space groups seems to reveal that these non-trivial twists can result in a change in the type of representation at fixed loci and hence in different K-theory at different points. We leave a full understanding of these twisted representation rings and a rigorous construction of the spectral sequence for non-symmporphic space groups to future work. 

In the introduction we explained an intuitive picture of the $\Z_2$ invariants in class AII that was put forward in \cite{Kruthoff2}. In particular, when crystal symmetries are present it was argued in \cite{Kruthoff2} that in two dimensions the vortex anti-vortex pair is stuck on fixed points whenever there is a rotation symmetry and stuck on fixed circles whenever there is a reflection symmetry. In string theory there is a analogous interpretation. Witten showed in \cite{Witten:1998cd} that there is a direct relation between the charges of $D$-branes on orbifolds and equivariant $K$-theory. Depending on what string theory is considered and whether there are involutions present, various versions of $K$-theory classify the corresponding $D$-brane charges. Moreover, at the orbifold singularity, say $\C^2/\Z_n$, only charge-$n$ D-branes can be peeled off, other charges are stuck on the singularity, branes with these charges are called fractional branes. In $K$-theory this means that only certain vector bundles see the singularity. This is similar to the fact that only $2$ vortex anti-vortex pairs can be moved away from a fixed point with $n$-fold rotation symmetry for topological insulators in class AII. Thus a single vortex anti-vortex pair is frozen on the fixed point. 

Although this frozen vortex picture gives the correct number of $\Z_2$ invariants for the example we computed, this interpretation is not immediately clear from our K-theory computations. One way to clarify this, is by using a localisation technique by Segal and Atiyah-Segal \cite{segalequivariant, PartIIAtiyahSegal}\footnote{We thank Gregory Moore for this suggestion.}. Originally, this is a result that applies to (untwisted) equivariant complex K-theory, $K_G(X)$, and uses the fact that $K_G(X)$ is an $R(G)$-module, with $R(G)$ the representation ring of $G$. Generalisations to other K-theories have also been mentioned in the literature, \cite{Distler:2009ri, NotesMoore}. For instance, in the latter reference, this technique has been applied to the three-dimensional diamond structure in class A, which has a non-trivial twist. For our purposes we would have to be generalise the localisation technique to cases in which a time-reversing operator is present as well, which we hope to pursue in future work.

\subsection*{Acknowledgement}
It is a pleasure to thank Gregory Moore, Peter Teichner, Bernardo Uribe and Jasper van Wezel for their engaging discussions. JK is supported by the Delta ITP consortium, a program of the Netherlands Organisation for Scientific Research (NWO) that is funded by the Dutch Ministry of Education, Culture and Science (OCW).

\appendix

\section{Appendix}\label{appendix}
\subsection{Freed \& Moore $K$-theory and twists}
\label{sec:Freed Moore twists}

In order to illustrate our approach to computing these $K$-theory groups, let us first rigorously define the notions used in the text.
While doing this, we also connect the technical mathematical language of Freed \& Moore \cite{Freed2013} to our more concrete setting.
To motivate this, first consider a $2+1$-dimensional square crystal and a finite classical symmetry group $G$ consisting of a time-reversing symmetry $T$ and a spatial symmetry $R$ of rotation by $\pi$.
To account for the fact that $R$ acts unitarily and $T$ acts antiunitarily, we define a homomorphism $\phi: G \to \mathbf{Z}_2$ by $\phi(R) = 1$ and $\phi(T) = -1$.
Although the classical group is $G =\mathbf{Z}_2 \times \mathbf{Z}_2$, we know that for fermions, we have $T^2 = R^2 = -1$ on the quantum level.
Hence we are not interested in modules over the group algebra of $G$, but in modules over a twisted group algebra.
To implement this fact mathematically, we twist the group $G$ by a group 2-cocycle $\tau \in Z^2(G,U(1))$, where $U(1)$ is the circle group seen as a $G$-module by $T e^{i \theta} = e^{-i \theta}$ and $R e^{i \theta} = e^{i \theta}$.
This cocycle is given by
\be
\label{eq:rotationtwist}
\t(T,T) = -1, \quad \t(R,R) = -1, \quad \t(T,R) = \t(R,T) = 1.
\ee
We extend this definition to a cocycle on all of $G$ by the cocycle relation and demanding that $\tau(g,1) = \tau(1,g) = 1$ for all $g \in G$.
On the quantum level of twisted representations, we want to impose equations such as $T \cdot T = \tau(T,T)$ instead of the equations holding in $G$.

To implement this in a more general setting, suppose that we are given a finite classical symmetry group $G$ consisting of point group symmetries and (possibly) time-reversal symmetry.
Let $\phi: G \to \mathbf{Z}_2$ be a homomorphism determining whether a group element acts unitarily or antiunitarily.
Consider a group $2$-cocycle $\tau \in Z^2(G,U(1)_\phi)$, where $U(1)_\phi$ is the $G$-module $g \cdot e^{i \theta} := e^{\phi(g)i \theta}$.
By the one-to-one correspondence between group extensions and group cohomology, the data of $\tau$ is equivalent to what is called a $\phi$-twisted extension in Freed \& Moore \cite{Freed2013}.
This is a group extension
\be
1 \to U(1) \to G^\tau \overset{\pi}{\to} G \to 1
\ee
such that $e^{i \theta} g = g e^{\phi(\pi(g))i \theta}$ for all $g \in G^\tau$.
In fact, two such $\phi$-twisted extensions are isomorphic if and only if the corresponding group 2-cocycles are cohomologous.
Therefore only the cohomology class $[\tau] \in H^2(G,U(1)_\phi)$ of the cocycle is relevant for the theory.

Now we can define how to twist representations of $G$ by $\phi$ and $\t$.
Indeed, a $(\phi,\t)$-projective action of $G$ is a map $\rho: G \to GL_\R(V)$ into the real linear automorphisms of a complex vector space $V$ such that $\rho(g)$ is complex linear if $\phi(g) = 1$, complex antilinear if $\phi(g) = -1$ and
\be
\rho(g) \rho(h) = \tau(g,h) \rho(gh).
\ee
Note in particular that if $\tau$ is nontrivial, $\rho$ is not a homomorphism of groups.
Via the correspondence between group cocycles $\tau$ and extensions, such projective actions are exactly the same as $(\phi,\tau)$-twisted representations in the sense of Freed and Moore \cite{Freed2013}.
These are defined as genuine homomorphisms $\rho^\tau: G^\tau \to GL_\R(V)$ into the real linear automorphisms of a complex vector space $V$ such that $\rho^\t(g)$ is complex linear if $\phi^\t(g) = 1$, complex antilinear if $\phi^\t(g) = -1$ and $\rho^\t(z)$ is just multiplication by $z$ if $z$ is in the circle subgroup $U(1) \subseteq G^\t$.
Therefore we will use $(\phi,\t)$-projective actions and $(\phi,\t)$-twisted representations interchangeably. 

The twist $\tau$ can also be used to twist the group algebra as follows.
We define the twisted group algebra $^\phi \C^{\tau} G$ to be the $2 \cdot \#G$-dimensional algebra over $\R$ generated by the symbols $x_g$ for every $g \in G$ and a formal imaginary unit $i$ with defining relations
\be
x_g x_h = \tau(g,h) x_{gh}, \quad i^{2} = -1, \quad x_g i = \phi(g) i x_g.
\ee
If no confusion can arise we usually just write $g$ for the symbol $x_g$.
Modules over the twisted group algebra are clearly equivalent to projective actions with cocycle $\tau$ and hence equivalent to $(\phi,\tau)$-twisted representations.

For example, suppose we consider time-reversal symmetry $T$ and $n$-fold rotation $R$ in a system of spinful fermions.
Then the symmetry group is $G = \Z_n \times \Z_2$ and $\phi$ is projection on the second factor. 
It can be shown using basic techniques in group cohomology that
\be
H^2(G,U(1)_\phi) = 
\begin{cases}
\Z_2^2 \text{ if } n \text{ is even,}
\\
\Z_2 \text{ if } n \text{ is odd.}
\end{cases}
\ee
For $n$ even, the two $\Z_2$'s correspond exactly to the choices of signs in $R^n = \pm 1$ and $T^2 = \pm 1$.
For example, if $n = 2$, a representative $\t$ of the cohomology class that assumes for both signs the negative is given in equation \eqref{eq:rotationtwist}.
For $n$ odd however, the sign of $R$ does not influence the isomorphism class of the twist.
This is not very surprising from a representation-theoretic perspective.
Indeed, if we redefine $S := -R$ in the group algebra $^\phi \C^{\tau} G$ then we get the group algebra with the twist chosen such that $S^n = -1$ and $T$ has the same square as before.
Note that this would not work for even $n$; we could have defined $S := iR$, but then $S$ would not commute with $T$.
However, if there would have been no time-reversal symmetry, this argument would have worked and the sign of $R^n$ does not matter for the cohomology class.
This simply resonates the fact that $H^2(\Z_n,U(1)) = 0$ in case of trivial $\phi$.
Conclusively, assuming that $R^n = -1$ in class A would not influence the classification of topological insulators and assuming that $R^n = 1$ in class AII would not influence the classification in case $n$ is odd.

The  group  cocycle $\t$ can  be  used  as  a twisting  for the Freed-Moore $K$-theory groups.
The $G$-equivariant $K$-theory of the Brillouin zone torus twisted by $\t$ then classifies topological phases protected by the twisted symmetry group $(G,\phi,\tau)$.
In the abstract language of Freed and Moore, to twist a $G$-space $X$ means that we consider the following $\phi$-twisted extension of the action groupoid (or orbifold) $X/ \hspace{-1.3mm}/G$.
The line bundle is picked trivial and the cocycle on $X/ \hspace{-1.3mm}/G$ is picked equal to the cocycle $\t$ at every $x \in X$, see also \cite[section 2]{gomi}.
Twisted equivariant bundles are then the bundle-theoretic analogue of $(\phi,\t)$-twisted representations of $G$ in the same sense that Atiyah \& Segal's complex equivariant vector bundles are the bundle-theoretic analogue of ordinary group representations of $G$.
More concretely, we define a $(\phi,\t)$-twisted equivariant vector bundle over a $G$-space $X$ to be a complex vector bundle $E$ over $X$ together with a family $\{\rho(g): E \to E: g \in G\}$ of maps such that 
\begin{itemize}
    \item[(i)] $\rho(g)$ covers the action of $g$ on the base space;
    \item[(ii)] $\rho(g)$ is complex linear if $\phi(g) = 1$ and conjugate linear if $\phi(g) = -1$;
    \item[(iii)] $\rho(g) \rho(h) = \tau(g,h) \rho(gh)$.
\end{itemize}
The $K$-theory classifying such bundles is simply the Grothendieck completion of the monoid of isomorphism classes of such bundles.
Written out in full, the resulting abelian group
\[
^\phi K^{\tau}_G(X)
\]
is called the $(\phi,\tau)$-twisted $G$-equivariant $K$-theory of $X$.
Twisted equivariant $K$-theory of this form can be expanded to a contravariant functor from a category of sufficiently nice $G$-spaces to the category of abelian groups.
By mimicking the technique of Segal \cite{segalequivariant}, it can be shown that this theory extends to a $\mathbf{Z}$-graded additive generalized equivariant cohomology theory in the sense of Bredon \cite{bredonequivariant}.
One can also show that the twisted equivariant $K$-theory above is equivalent to the $K$-theory of Freed \& Moore (with our specific form of the twist $\tau$) under the correspondences described above c.f.\ \cite{gomi}.
The fact that Freed \& Moore's $K$-theory satisfies the axioms desired for a cohomology theory of this kind also follows from \cite{gomi}. Hence we can use the equivariant form of the usual cohomology axioms (suspension axiom, homotopy invariance, etc.) freely.
The associated reduced cohomology theory is called reduced twisted equivariant $K$-theory.
For a pointed $G$-space $(X,x)$, where $x \in X$ is a point that is completely fixed under the action, the reduced $K$-theory is defined as the kernel of the map given by restricting to the fiber over the base point:
\be
^\phi \tilde{K}^{\t + p}_G(X) := \ker\left(^\phi K^{\t + p}_G(X) \to ^\phi K^{\t + p}_G(x)\right).
\ee
If $Y \subseteq X$ is a subspace closed under the $G$-action, then the relative twisted equivariant $K$-theory of the pair $(X,Y)$ is defined as
\be
^\phi K^{\t + p}_G(X,Y) := {}^\phi \tilde{K}^{\t + p}_G(X/Y)  
\ee
It should also be noted that that the $K$-theory twisted by a cocycle $\t'$ cohomologous to $\t$ is isomorphic to the $K$-theory twisted by $\t$.

Finally we remark on how to formally construct the necessary $\t$ for classifying class AII crystalline topological insulator.
We do this by considering the double cover group sketched in Section \ref{Othersymm}.
Before we can do this, we first have to sketch the relation between double cover groups and twists $\t$.
In order to show this, suppose we have a group of the form $G = G_0 \times \Z_2^T$ with $\phi$ projection onto the second factor and $G_0\hookrightarrow O(d)$ a point group.
As described using group extensions in Section \ref{Othersymm}, we start with the class in $H^2(O(d), \mathbf{Z}_2)$ corresponding to the $\operatorname{Pin}_-$-double cover of $O(d)$.
Restricting the class to the subgroup $G_0 \subseteq O(d)$ then gives a class $[\tau_1] \in H^2(G_0,\Z_2)$, which exactly corresponds to the double cover group $\widehat{G}_0$.
To get the right square of time-reversal, we then take the class $[\tau_2] \in H^2(\Z_2^T,\Z_2) = \Z_2$ to be trivial in class AI and nontrivial in class AII.
Next, in order to construct the desired total double cover group which covers $G$ instead of $G_0$, we now combine $[\tau_1]$ and $[\tau_2]$.
To do this we pull the classes back along the two projection maps $p_1: G \to G_0$ and $p_2: G \to \Z_2^T$ and then take their product:
\be
[\tau'] = p_1^*([\tau_1]) \cdot p_2^*([\tau_2]) \in H^2(G,\Z_2).
\ee
Note that this product is not the cup product, but simply the product on the level of the coefficients $\Z_2$.
Finally, we get our desired twist $[\tau] \in H^2(G, U(1)_\phi)$ by extending the coefficients to $U(1)$.
More precisely, we consider the map $H^2(G, \Z_2) \to H^2(G, U(1)_\phi)$ induced by the $G$-module injection $\Z_2 \hookrightarrow U(1)_\phi$.

More generally, we could have taken group cocycles $\tau \in Z^2(G, C(X,U(1)_\phi))$ that vary over the Brillouin zone $X$ to account for nonsymmorphic crystal structures. 
These more general twists can be used as a $\phi$-twisted extension of the action groupoid $X / \hspace{-1.3mm}/G$ in the Freed-Moore framework by picking the line bundle to be trivial and the cocycle to be equal to $\tau$, now varying over space.
This construction results in a well-defined $K$-theory group that classifies nonsymmorphic crystalline topological insulators.
However, we can no longer define the relative $K$-theory of $(X,Y)$ to be the reduced $K$-theory of the quotient if $\t$ varies along $X^{p-1}$.
For a definition of relative $K$-theory that holds in a more general setting, see \cite{gomi}.

\subsection{Higher twisted representation rings and the $K$-theory of a point}
\label{sec: twisted reprings}

Assume we are given a finite group $G$, a homomorphism $\phi: G \to \mathbf{Z}_2$ and a group 2-cocycle with values in the $G$-module $g \cdot e^{i\theta} = e^{i \phi(g) \theta}$.
The basic building blocks of $K$-theory from which spectral sequences can be built are the higher degree $K$-theory groups of a point $^\phi K^{\tau - q}_G(\pt)$.
It follows from the last section that for $q = 0$, the group $^\phi K^\tau_G(\pt)$ is the Grothendieck completion of the monoid of isomorphism classes of modules over the group algebra $^\phi \mathbf{C}^\tau G$ twisted by $\t$.
This is what is called the twisted representation ring of degree $q = 0$ in the main text.
Remark that due to the twist $\tau$, these abelian groups do not have a ring structure, but we will nevertheless refer to them as twisted representation rings in analogy with the nontwisted case.

In higher degrees, the twisted equivariant K-theory of a point has a similar concrete description in terms of representation theory and Clifford algebras.
This is expressed by using a generalized version of the Atiyah-Bott-Shapiro isomorphism \cite{abs}. 
Heuristically, this theorem asserts that going down a degree in $K$-theory (i.e. taking a suspension) corresponds to adding a Clifford algebra element on the algebraic level, at least for the $K$-theory of a point.
More explicitly, let $Cl_{p,q}$ for the real Clifford algebra of signature $(p,q)$.
Then $Cl_{p,q}$ is an algebra over the real numbers with has a natural $\Z_2$-grading such that the standard generators $\gamma_1, \dots, \gamma_{p+q} \in Cl_{p,q}$ are odd.
If we then take group elements to be even, we can give the tensor product $^\phi \mathbf{C}^\tau G \otimes_\R Cl_{0,q}$ the structure of a $\Z_2$-graded algebra as well.
The $(\phi,\t)$-twisted equivariant $K$-theory of a point can then be described by the isomorphism
\begin{align}
\label{eq:higherreprings}
   {}^\phi K^{\tau-q}_G(\pt)\cong {}^\phi R^{\tau - q}(G) : = 
    \frac{\{ \mathbf{Z}_2 \text{-graded modules over }^\phi \mathbf{C}^\tau G \otimes_\R Cl_{0,q} \}}
    {\{\text{modules that extend to a }^\phi \mathbf{C}^\tau G \otimes_\R Cl_{1,q} \text{-module}\}},
\end{align}
and we call ${}^\phi R^{\tau - q}(G)$ the $(\phi,\tau)$-twisted (higher) representation ring of $G$ in degree $-q$.
This fact follows from the discussion in \cite[\S 3.5]{gomi}.
This way of computing the complex and real equivariant $K$-theory of a point has been known for a long time, see the final section of \cite{atiyahsegal}.
Also see Donovan-Karoubi \cite[\S 6.15]{karoubi} for a precise version of such a statement for a certain type of twisted equivariant $K$-theory of more general spaces.
We again stress that the twisted higher representation rings as defined above are not rings themselves, because the tensor product of two $(\phi,\t)$-twisted representations is a $(\phi,2\tau)$-twisted representation.
However, they are modules over the $(\phi,1)$-twisted representation ring.
We could in theory make them into genuine rings by summing over all possible twists $\t$, but this would not be a very natural thing to do from the perspective of physics.
Namely, using the language of Section \ref{Othersymm}, the product of two fermionic representations will be a bosonic one.

To illustrate the definition of the twisted representation ring, we provide a few examples.
First of all, if $\phi$ and $\t$ are trivial, it can be shown that the higher representation rings are equal to the representation ring of the group in even degree and vanish in odd degree, i.e. it gives the complex equivariant $K$-theory of a point as expected. 
For certain simple groups like the time-reversal group $G = \mathbf{Z}_2 = \{1,T\}$ (with $\phi(T) = -1$), the twisted representation rings are also readily computed using basic Clifford algebra theory. They are simply the real (respectively quaternionic) $K$-theory of a point for a trivial (respectively nontrivial) twist $\t$. This results in the relevant representation rings for both class AI with trivial twist $\t_0$ and class AII with twist $\t_1$, as was summarized in Table \ref{TwistedRing}. 
If we just take the real group algebra instead of the twisted group algebra, we get the equivariant $KR$-theory of a point as can be shown by comparing with the results in \cite{atiyahsegal}.
It is also worth noting that the higher twisted representation ring defined by \eqref{eq:higherreprings} indeed equals the Grothendieck group of the monoid consisting of isomorphism classes of $(\phi,\t)$-twisted representations of $G$ in case $q =0$.

The higher representation rings of the twisted group algebra in general can be computed by decomposing the algebra into a direct sum of matrix rings over $\mathbf{R}, \mathbf{C}$ and $\mathbf{H}$.
Since the twisted group algebra is semisimple, this can always be done.
If the number of real, complex and quaternionic matrix rings occurring in this decomposition are $n_\mathbf{R}, n_\mathbf{C}$ and $n_\mathbf{C}$ respectively, then
\be
^\phi R^{\t - q}(G) = K^{-q}(\pt)^{n_{\mathbf{C}}} \oplus KR^{-q}(\pt)^{n_{\mathbf{R}}} \oplus KR^{-q-4}(\pt)^{n_{\mathbf{H}}}.
\ee
This follows because the representation rings preserve direct sums and are independent of Morita equivalence.
Therefore, to determine the representation rings, we only need to make an analogous twisted representation theory to the theory of real representations of finite groups.
Such a theory already exists in the physics literature and is described in Section \ref{Othersymm}.

\subsection{Construction of the spectral sequence} \label{sec:spectral sequence}

In this section some details on the existence of the spectral sequence will be outlined.
For an introduction to the theory used in this section and throughout the paper, such as spectral sequences in general and the Atiyah-Hirzebruch spectral sequence for (nonequivariant) cohomology theories such as ordinary $K$-theory, we refer the reader to \cite{hilton1971general}.
The construction of the spectral sequence for the type of $K$-theory we are concerned with can be done using standard methods already described for general equivariant cohomology theories by Bredon \cite{bredonequivariant}.

As before, let $G$ be a finite group, $\phi: G \to \mathbf{Z}_2$ a homomorphism and $\tau \in Z^2(G,U(1)_\phi)$ a group cocycle.
We now start with a finite $G$-CW complex $X^0 \subseteq \cdots \subseteq X^d = X$ and construct the Atiyah-Hirzebruch spectral sequence with respect to this $G$-CW-structure.
In order to make sure that the reduced $K$-theory of $X$ is defined, we have to assume that there is at least one point $0 \in X^0$ that is fixed by the whole group $G$.
Note that this is always the case in practice; if $X$ is the Brillouin zone torus then the point $k = 0$ is fixed by all symmetries.
As stated in \cite{bredonequivariant}, the first page $E_1^{p,-q}$ of the spectral sequence is given by the relative $K$-theory groups
\be
E_1^{p,-q} = {}^\phi K_G^{\tau + p - q}(X^p,X^{p-1}).
\ee  
Let us now show that this is the group of Bredon equivariant cochains with a particular coefficient functor. Indeed first note that since $\t$ is constant in space, the relative $K$-theory equals reduced $K$-theory of the quotient:
\begin{align}
    {}^\phi{K}{^{\tau + p-q}_G}(X^p,X^{p-1}) &= {}^\phi{\widetilde{K}}{^{\tau + p-q}_G}(X^p/X^{p-1}) 
    \\
    &\cong {}^\phi{\widetilde{K}}{^{\tau + p-q}_G}\left(\bigvee_{\sigma} \bigvee_{g H_\sigma} S^p_{gH_{\sigma}}\right).
\end{align}
Here the first wedge product is over all equivariant $p$-cells $\sigma$ of $X$, which look like $S^p_\sigma \times G/H_{\sigma}$, where $H_\sigma \subseteq G$ is the stabilizer of the cell $\sigma$.
The second wedge product is over all ordinary cells $S^p_{g H_\sigma}$ contained in the equivariant cell $\sigma$, i.e. over all cosets $gH_\sigma$ of the stabilizer group.
The appearance of the wedge sums is a consequence of the quotient $X^p/X^{p-1}$. For example, if $X = S^2$ is the sphere of figure \ref{fig:CW structure} and $p = 2$, then $X^p/X^{p-1}$ is the space that results from pinching $\ell$ and $T \ell$ to a point, i.e. $S^2 \vee S^2$.

Now by additivity and the suspension axiom,
\begin{align}
    {}^\phi{K}{^{\tau + p-q}_G}(X^p,X^{p-1}) &= {}^\phi{\widetilde{K}}{^{\tau + p-q}_G}\left(\bigvee_{\sigma} \bigvee_{g H_\sigma} S^p_{gH_{\sigma}}\right)
    \\
    &\cong \bigoplus_{\sigma \,\, p\text{-cell}} {}^\phi{\widetilde{K}}{^{\tau + p-q}_G}\left(\bigvee_{g H_\sigma} S^p_{gH_{\sigma}}\right)
    \\
    &= \bigoplus_{\sigma \,\, p\text{-cell}} {}^\phi{\widetilde{K}}{^{\tau + p-q}_G}(\Sigma^p(G/H_{\sigma} \sqcup \pt))
    \\
    &\cong \bigoplus_{\sigma \,\, \,\, p\text{-cell}} {}^\phi{\widetilde{K}}{^{\tau - q}_G}(G/H_{\sigma}  \sqcup \pt)
    \\
    &\cong \bigoplus_{\sigma \,\, p\text{-cell}} {}^\phi{K}_G^{\tau - q}(G/H_{\sigma}).
\end{align}
Here $\Sigma^p Y$ denotes the $p$th reduced suspension of the pointed $G$-space $Y$ and $\sqcup \pt$ is the disjoint union with an extra added basepoint.
Similarly to nontwisted equivariant $K$-theory we can then make use of the isomorphism
\begin{align}
    {}^\phi{K}_G^{\tau - q}(G/H)
    &\cong {}^\phi{K}_H^{\tau - q}(\pt)
\end{align}
induced by restricting bundles over $G/H$ to the fiber lying over the trivial coset $H$.
Note that on the right hand side, we have to restrict the twisting data $\phi,\tau$ to $H$, but this is omitted in the notation.

The first page of the spectral sequence can now be rewritten as a group of Bredon equivariant cochains
\begin{align}
    \bigoplus_{\sigma \,\, p\text{-cell}} {}^\phi{K}_G^{\tau - q}(G/H_{\sigma}) &\cong
    \bigoplus_{\sigma \,\, p\text{-cell}} {}^\phi{K}_{H_{\sigma}}^{\tau - q}(\pt)
    \\
    &\cong \bigoplus_{\sigma \,\, p\text{-cell}} {}^\phi{R}{^{\tau - q}}(H_{\sigma})
    \\
    &\cong C^p_G(X,{}^\phi{\mathcal{R}}{^{\tau - q}_G}),
\end{align}
where the coefficients ${}^\phi{\mathcal{R}}{^{\tau - q}_G}$ form a functor from the orbit category of $G$ to the category of abelian groups. 
Topologically, this functor is just the restriction of the twisted equivariant $K$-theory functor to the orbit category, but an algebraic desciption is more enlightening.
It sends the orbit space $G/G_{\s}$ to the twisted representation ring ${}^\phi R{^{\tau - q}}(G_{\s})$ defined in the previous section:
\be
{}^\phi{\mathcal{R}}{^{\tau - q}_G}(G/G_{\s}) =  {}^\phi R{^{\tau - q}}(G_{\s}).
\ee 
Thus although our cochains appear to have coefficients in a functor, once evaluated for specific cells in the CW complex of $X$, the coefficients are just the degree $-q$ twisted representation ring of the stabilizer group of that cell. 
However, its action on morphisms is more complicated to describe algebraically.
Quotient maps $G/H \to G/K$ are sent to restrictions of representations ${}^\phi R^{\tau - q}(K) \to {}^\phi R^{\tau - q}(H)$ as expected, but conjugation maps $G/H \to G/gHg^{-1}$ can yield nontrivial results similar to the action of $T$ as described around equation \eqref{nontrivialTaction}.
As stated in the main text, it can be shown that the first differential is precisely the cellular Bredon differential $d$
\be
(d f)(\sigma) = \sum_{\mu \in C^p(X)} [\mu : \sigma] f(\mu)|_{G_\sigma},
\ee
see for example Bredon's work \cite{bredonequivariant}.

Summarizing, there exists a spectral sequence $E^{p,-q}_r$ associated to a finite pointed $G$-CW-complex $X$ converging to twisted equivariant $K$-theory:
\be
E^{p,-q}_r \implies {}^\phi K^{\t + p - q}_G (X)
\ee
such that the second page $E^{p,q}_2$ is Bredon cohomology of degree $p$ with coefficient functor ${}^\phi{\mathcal{R}}^{\tau - q}_G$.
To derive explicit computational tools from this fact, recall from the basic theory of spectral sequences that this means that there is a filtration $F^p$ of $^\phi K^{\t}_G (X)$
\be
0 = F^{d+1} \subseteq F^{d} \subseteq \dots \subseteq F^1 \subseteq F^0 = {}^\phi K^{\t}_G (X)
\ee
such that the final page $E^{p,-p}_\infty$ forms the associated graded space, i.e.
\be
E^{p,-p}_\infty \cong \frac{F^{p}}{F^{p+1}}.
\ee
If $d = 2$, this results in the short exact sequences \eqref{eq: exact sequences1} and \eqref{eq: exact sequences2}, where $F = F^1$.

For computations for nonsymmorphic crystals, we would need to generalize these methods to include twisting data for $K$-theory that varies over $X$.
Since $\tau$ is no longer constant in this setting, most arguments given in this section fail.
First of all, the relative $K$-theory of $(X^p,X^{p-1})$ no longer seems to be equal to the $K$-theory of the quotient $X^p/X^{p-1}$ if $\t$ varies along $X^{p-1}$, so we have to preserve the information of the values of $\t$ over $X^{p-1}$.
Moreover, it is not clear how to compute the twisted equivariant $K$-theory of a sphere for nonconstant twist, since there is no obvious generalization of the isomorphism \eqref{eq:higherreprings} here.
Ignoring these mathematical difficulties for the moment and assuming we have some kind of spectral sequence, we should at least arrive at a point where the coefficient functors ${}^\phi\mathcal{R}^{\t-q}_G$ are no longer constant.
In particular, a Bredon cochain of say degree $1$ could map different points of a single $1$-cell into different twisted representation rings.
The second page is no longer ordinary Bredon cohomology and therefore it is not clear how to generalize this section to that setting.

\bibliography{Ktheory.bib}
\bibliographystyle{Ktheory}

\end{document}